\documentclass[twocolumn]{aastex631}

\usepackage{natbib}
\usepackage{amssymb}
\usepackage{amsmath}
\mathchardef\mhyphen="2D % Define a "math hyphen"
\usepackage{verbatim}
\usepackage{indentfirst}
\usepackage{xcolor}
\usepackage{graphicx}
\usepackage{hyperref}
\usepackage{makecell}

\newcommand{\Rnum}[1]{\uppercase\expandafter{\romannumeral #1\relax}}

\newcommand{\ha}{{\rm H\ensuremath{\alpha}}}

\newcommand{\mbh}{{\ensuremath{M_{\bullet}}}}

\newcommand{\mstar}{{\ensuremath{M_\star}}}

\newcommand{\msun}{{\ensuremath{{M_\odot}}}}

\newcommand{\valerrud}[3]{${#1}^{+#2}_{-#3}$}
\newcommand{\verrud}[3]{{#1}^{+#2}_{-#3}}

\newcommand{\condp}[2]{\ensuremath{\left(#1\middle|#2\right)}}
\newcommand{\clog}{\ensuremath{\log_{10}}}
\newcommand{\fha}{\ensuremath{F_\mathrm{b\ha{}}}}

\newcommand{\mailto}[1]{{\normalfont\href{mailto:#1}{#1}}}

\newcommand{\snu}{\affil{Department of Physics \& Astronomy, Seoul National University, Seoul 08826, Republic of Korea; \mailto{jhwoo@snu.ac.kr}}} 
\begin{document}

\title{Constraining the Low-Mass End of the Black Hole Mass Function and the Active Fraction of the Intermediate-mass Black Holes}

\author[0000-0003-2010-8521]{Hojin Cho} \snu
\author[0000-0002-8055-5465]{Jong-Hak Woo} \snu

% \correspondingauthor{Jong-Hak Woo}\email{jhwoo@snu.ac.kr}

%%%%%%%%%%%%%%%%%%
% Abstract       %
%%%%%%%%%%%%%%%%%%

\begin{abstract}
We investigate the black hole mass function (BHMF) and the Eddington ratio distribution function (ERDF), focusing on the intermediate-mass black holes (IMBHs) with masses down to $M_{\bullet}\sim10^4 M_\odot$. Based on the active galactic nuclei (AGNs) with a detected broad H$\alpha$ emission line, we construct a sample of 14,242 AGNs at redshift $z<0.35$, including 243 IMBHs with $M_{\bullet}<10^6 M_\odot$. By jointly modeling the BHMF and ERDF via the maximum posterior estimation, we find that the BHMF peaks at $\sim$$10^{6} M_\odot$ and exhibits a relatively constant value of $10^{-4}\,\mathrm{Mpc^{-3}\,dex^{-1}}$ at the low-mass end. By comparing the derived BHMF of type 1 AGNs with the galaxy mass function based on the updated black hole mass--host galaxy stellar mass relation, we derive the active fraction. We also determine the active fraction for all AGNs using the upper and lower limit of the type 1 fraction. The active fraction decreases from 15\%--40\% for massive galaxies ($\mstar>10^{10}\msun$) to lower than $\sim$2\% for dwarf galaxies with $\mstar\sim10^8\msun$. These results suggest that the black hole occupation fraction is expected to be $\sim$50\% for low-mass galaxies ($\mstar\sim10^{8.5}$--$10^9\msun$) if the duty cycle is similar IMBHs and supermassive black holes.
\end{abstract}

%%%%%%%%%%%%%%%%%%
% Introduction   %
%%%%%%%%%%%%%%%%%%

\section{Introduction}
The origin and evolution of supermassive black holes (SMBHs) are not well understood yet, despite their significance in galaxy evolution. The presence of SMBHs in almost all massive galaxies, as well as the correlation with the host galaxies, suggests the coevolution of black holes and galaxies \citep[e.g.,][]{Rees1984, Kormendy&Richstone1995, Ferrarese&Merritt00, Gebhardt+00, Kormendy&Ho13, Woo+2013, Heckman&Best2014}.

At the low-mass end of the mass spectrum are the intermediate-mass black holes (IMBHs), which are typically defined as black holes with mass of $10^2\msun\leq\mbh<10^6\msun$. IMBHs are considered to hold a key to understanding the origin of SMBHs, as various theoretical models predict different mass functions and occupation fractions of IMBHs in dwarf galaxies even in the local Universe (see \citealt{Greene+20} for a review). For instance, the light-seed scenario argues that SMBHs originated from the remnants of Population III (Pop III) stars with a typical mass of $\sim$100\msun \citep[e.g.,][]{Haiman&Loeb2001, Madau&Rees2001, Heger+2003, Volonteri+03}. Since Pop III stars were prevalent in the early Universe, most of the galaxies in the current Universe should harbor SMBHs at their centers if SMBHs were formed by the light-seed channel. However, the presence of black holes heavier than $10^{10}\msun$ at $z\sim 6.3$ \citep{Wu+15} poses challenges for the light-seed scenario, as growth at super-Eddington rates would be necessary in a relatively short timescale. In contrast, the heavy-seed scenario suggests that the rapid collapse of halo gas in the early Universe formed IMBHs of $\mbh\sim10^5$--$10^6$\msun \citep[e.g.,][]{Loeb&Rasio1994, Bromm&Loeb2003, Begelman+2006, Lodato&Natarajan2006}. In this case, SMBHs are not as prevalent in dwarf galaxies as the light-seed scenario predicts \citep[e.g.,][]{Bellovary+2019}. However, this channel requires a mechanism that prevents the fragmentation of the gas \citep[see][and references therein]{Inayoshi+2020a}.

Thus, it is important to investigate the black hole mass function (BHMF) and the occupation fraction to understand the origin of SMBHs. Previous constraints on the BHMF, however, have been insufficient for testing the black hole seed scenarios. One approach is to infer the BHMF from the mass function (or anything equivalent, such as the luminosity function (LF)) of galaxies. \citet{Shankar+2004}, for instance, derived local BHMF down to $10^6\msun$ by convolving the local galaxy LF with the $\mbh$--$L_\mathrm{sph}$ relation \citep{McLure&Dunlop2002}, implicitly assuming that all galaxies harbor a black hole at their center. \citet{Gallo&Sesana19} advanced this approach further by accounting for the occupation fraction in the galaxies observed in the local Universe \citep{Miller+15}. While this approach is useful, it relies on the choice of the scaling relation of the black hole mass against the host galaxy property, which is not well constrained for the low-mass black holes \citep[e.g.,][]{Kormendy&Ho13, Reines&Volonteri2015, Woo+19a}. Furthermore, this approach cannot be employed to derive the occupation fraction, which is required for computing the BHMF.

The virial mass estimates of active galactic nuclei (AGNs) can provide constraints on the BHMF since the number of AGNs is the lower limit of the number of black holes. For instance, \citet{Kelly+2009} demonstrated a Bayesian method to jointly model the BHMF and Eddington ratio distribution function (ERDF) as Gaussian mixtures. By applying the method to the 87 quasars in the Bright Quasar Survey \citep{Schmidt&Green1983}, including 16 with reverberation mass estimates, they constrained the local ($z<0.5$) BHMF down to $\mbh\sim10^8\msun$. \citet{Kelly&Shen2013} applied the same method to the quasars in the Sloan Digital Sky Survey \citep[SDSS;][]{York+2000} Data Release 7 (DR7), resulting in a similar BHMF for $\mbh{}>10^8\msun$. \citet{Schulze&Wisotzki2010} utilized AGNs in the Hamburg/ESO Survey \citep{Wisotzki+2000} and constrained the BHMF down to $10^6\msun$ based on a similar parametric approach assuming various functional forms of the BHMF and ERDF. \citet{Greene&Ho2007} were one of the first to provide BHMF below $10^6\msun$ threshold by selecting AGNs in SDSS and measuring their \mbh{} based on the broad \ha{} emission. However, the BHMF and LF they derived with the $1/V_\mathrm{max}$ method \citep{Schmidt1968} were substantially smaller than those in other studies, including the aforementioned ones, which may suggest that the selection completeness they used was overestimated. Ultimately, these ``BHMFs'' only trace the mass function of active black holes; thus, they cannot be directly compared with the galaxy mass function to derive the occupation fraction without assumptions on the black hole activity, such as the duty cycle.

In this work, we present an improved estimate of the BHMF and ERDF based on the broad \ha{}-selected AGNs in SDSS DR7 by modeling the sample completeness more accurately than in previous studies. In Section~2, we define the sample and quantify the selection bias in our sample. Section~3 describes the mathematical details of our modeling. The results of our modeling are presented in Section~4, followed by a discussion in Section~5.
% Cosmology
Throughout this paper, we adopt a flat $\Lambda$CDM cosmology with $H_0 = 72\,{\rm km\, s^{-1}\, Mpc^{-1}}$ and $\Omega_{m} = 0.3$.

%%%%%%%%%%%%%%%%%%
% Sample         %
%%%%%%%%%%%%%%%%%%

\section{The Sample}\label{s:sample}
The sample of AGNs we used for measuring the BHMF is based on SDSS \citep{York+2000}, which is a comprehensive optical survey of photometry and spectroscopy. Its seventh data release covers an area of over 10,000~deg$^2$, with photometry of 357 million objects and 1.6 million spectra. The details of the survey were described by \citet{Abazajian+2009}.

SDSS used the \emph{ugriz} filter system \citep{Fukugita+1996} for its photometry. Based on their photometric measurements, they compiled two sets of spectroscopic targets covering 8032~deg$^2$. The main galaxy sample \citep{Strauss+2002} consists of objects with $r$-band \citet{Petrosian1976} magnitudes brighter than 17.7~mag ($r_\mathrm{Petro}\leq17.7$) and not a point source ($r_\mathrm{PSF}-r_\mathrm{model}\geq 0.3$). Similarly, the quasar sample \citep{Richards+2002} consists of objects with $i$-band PSF magnitudes brighter than 19.1~mag ($i_\mathrm{PSF}<19.1$). Furthermore, objects that would saturate the spectrograph were rejected ($i_\mathrm{fiber}<14.5$ or $r_\mathrm{fiber}<15$).

\citet{Liu+2019} carried out an exhaustive search of AGNs within the galaxy and quasar samples that exhibit broad \ha{} emission lines in their spectra. After removing the stellar continuum and modeled broad \ha{} lines, they selected AGNs based on the following criteria: (1) $p$-value smaller than 0.05 from the $F$-test for the broad component fitting, (2) broad \ha{} flux larger than $10^{-16}\,\mathrm{erg\,s^{-1}\,cm^{-2}}$, (3) broad \ha{} signal-to-noise ratio larger than 5, (4) flux density at the peak of broad \ha{} larger than twice the continuum-subtracted rms, and (5) an FWHM of the broad \ha{} line profile larger than that of narrow lines. The resulting sample consists of 14,584 broad-line AGNs within $z<0.35$, which corresponds to the redshift limit for \ha{} in the SDSS spectrograph, with their broad \ha{} luminosity spanning $10^{38.5}$--$10^{44.3}\,\mathrm{erg\,s^{-1}}$.

\begin{figure*}[ht!]
\centering
\includegraphics[width=0.9\textwidth]{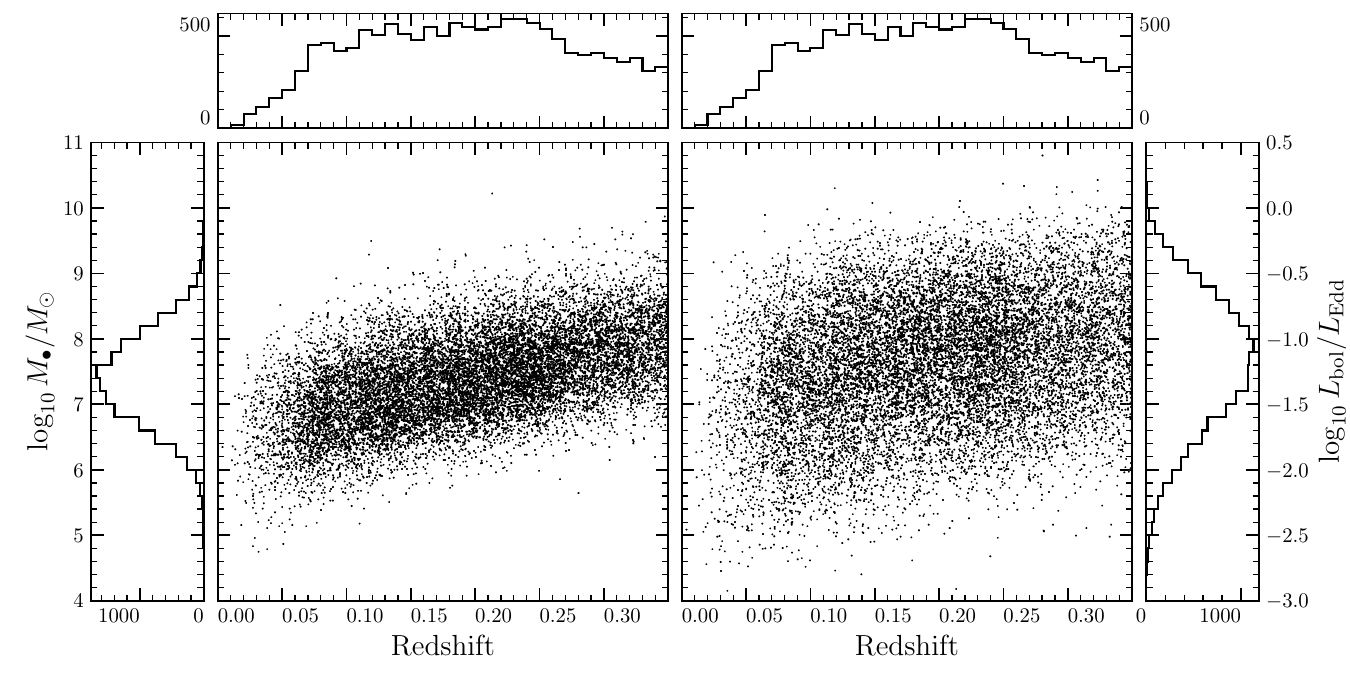}
\caption{Distributions of the mass (left) and the Eddington ratio (right) as a function of redshift, along with histograms of the redshift, mass, and Eddington ratio. 
\label{fig:sample}}
\end{figure*}

Based on the sample and measurements by \citet{Liu+2019}, we constructed our sample as follows. First, we eliminated any object with broad \ha{} flux larger than $7.5\times10^{-13}\,\mathrm{erg\,s^{-1}\,cm^{-2}}$ as such objects on average would be brighter than the saturation limit of the survey (see Section~\ref{ss:sed}). Then, we calculated broad \ha{} luminosity based on our cosmology and rejected objects with $L_\ha{}<10^{39}\,\mathrm{erg\,s^{-1}}$, as the number of objects below it was too small to produce meaningful statistics. We also estimated the black hole mass and the Eddington ratio from the luminosity and FWHM of broad \ha{} based on the single-epoch mass estimator and \ha{} bolometric correction of $L_\mathrm{bol} = \left(190\pm10\right)\times L_\ha{}$ provided by \citet{Cho+2023} and removed any object with Eddington ratio smaller than $10^{-3}$. The resulting sample consists of 14,242 AGNs, with their mass, luminosity, and Eddington ratio spanning $4.5<\clog\mbh/\msun<10.5$, $39<\clog L_\ha{}/\mathrm{erg\,s^{-1}}<44.5$, and $-3<\clog L_\mathrm{bol}/L_\mathrm{Edd}<0.5$, respectively. Note that our sample includes 243 active IMBHs with $\mbh<10^{6}\msun$. We present the distribution of the redshift, mass, and Eddington ratio of our sample in Figure~\ref{fig:sample}.

To correctly estimate the underlying population of AGNs, especially for the low-mass and low-luminosity end of the spectrum, the detection probability of each object in a sample, or \emph{the selection function}, needs to be determined as accurately as possible. This probability then can be used to estimate the number of AGNs that are not included in the sample because of the selection procedure. Our sample was constructed based on two distinct samples (i.e., SDSS quasar and main galaxy samples), and the selection process utilized the measurements of various optical bands, which could be contaminated by the stellar emissions of the host galaxy. While we cover a wide range of mass and luminosity by combining two distinct AGN samples, it is challenging to correctly model the selection function of our sample, which was selected based on the broad \ha{} emission lines. In contrast, host galaxy correction is not required for AGN samples selected from X-ray surveys, although X-ray binaries can be a challenge to properly constructing an AGN sample, particularly at low luminosity \citep[e.g.,][]{Gallo+2008, Baumgartner+2013}.

We derive the selection function of our sample as follows. In Section~\ref{ss:sed}, we construct the spectral energy distribution (SED) of pure AGNs without host galaxy contamination. Based on this, we derive the saturation limit of the AGNs in terms of \ha{} flux. Then, we model the distribution of the host galaxy fraction in Section~\ref{ss:selfunc}. Finally, we derive the selection function considering the $r$-band flux from both the AGN and the host galaxy.

\subsection{Spectral Energy Distribution of Quasars}\label{ss:sed}
The selection process of SDSS involves magnitudes of broadband photometry, which cannot separate the AGN from its host. To construct the selection probability based on the AGN luminosity only, the flux of broad \ha{} emission lines is a more robust proxy, as its distinctive feature can easily be separated from the host galaxy spectra. In this subsection, we investigate the correlation between $r$- or $i$-band flux and the broad \ha{} flux of AGNs, which we simply refer to as the SED of AGNs hereafter.

First, we compiled a list of objects with little to no host contamination. We defined the quasar subset of our sample, consisting of 7826 objects, to be the AGNs that were classified as quasar candidates but not as galaxy candidates based on the photometric selection criteria SDSS used \citep{Richards+2002, Strauss+2002}. However, a significant fraction of objects in this subset, typically fainter than 17.7 mag in the $r$ band, appear to have Petrosian magnitudes brighter than PSF magnitudes by more than 0.3 mag, indicating that their flux is dominated by the extended host galaxy. We note that the $r$ magnitude of 17.7 coincides with the magnitude limit of the galaxy sample of SDSS. Hence, we removed all objects with $r_\mathrm{Petro}>17.7$ and $r_\mathrm{PSF}-r_\mathrm{model}>0.1$. The latter condition is stricter than the corresponding selection criterion of the galaxy sample by \citet{Strauss+2002} in order to remove all potentially extended sources. The final subset of \emph{pure} AGNs consists of 1462 AGN-dominated objects.

\begin{figure*}[ht!]
\centering
\includegraphics[width=0.9\textwidth]{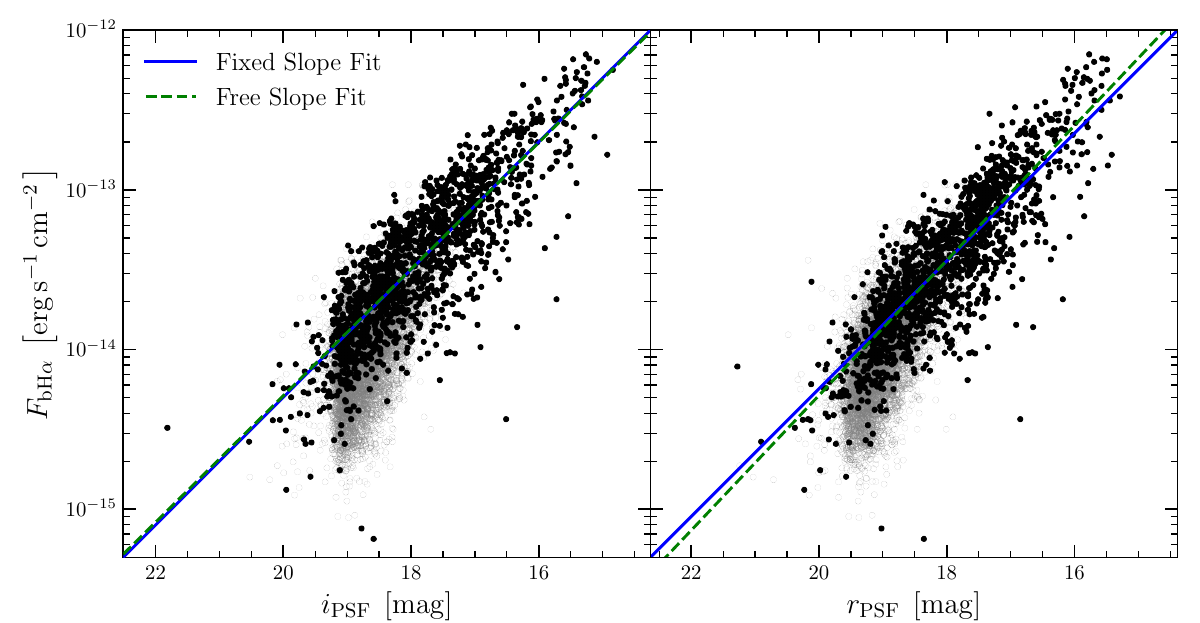}
\caption{Estimating the typical SED for the AGNs in the quasar subset. Black points were used for the fitting as described in Section~\ref{ss:sed}, while gray points show the entirety of the objects that were flagged by the SDSS photometric pipeline as quasars but not as galaxies. The best-fit relations, fixing the slope to unity, are given as solid lines. The linear fits without fixing slope are shown with dashed lines for comparison.
\label{fig:sed}}
\end{figure*}

To simplify the problem, we assumed that the shape of the SED does not depend on the luminosity of the AGN, i.e., the ratio of the $r$- or $i$-band flux to the broad \ha{} flux is the same for all AGNs. This is motivated by the \citet{Shakura&Sunyaev1973} disk model, where the SED of the accretion disk follows a power law $F_\nu \sim \nu^{1/3}$ for wavelengths $\lambda\gg10^3\mathrm{\AA}$ for a wide range of accretion rates. Observationally, \citet{Richards+2001} demonstrated that the $r-i$ color of SDSS quasars is consistent across different magnitudes except for the redshift effect. Furthermore, the luminosity of the broad \ha{} emission line is proportional to the optical continuum luminosity over a wide range of luminosities \citep[e.g.,][]{Yee1980, Shen&Liu12, Jun+15, Cho+2023}. Given the narrow range of redshifts of AGNs in our sample, we expect the ratio of $r$- or $i$-band flux to $\ha{}$ flux to be similar among the AGNs. Note that more distant AGNs may have different $r-i$ colors owing to the 3000\AA{} bump \citep{Grandi1982} or the Lyman-$\alpha$ forest affecting the magnitudes, which should be taken into account when modeling the selection function of the optically selected AGNs at high $z$.

Under the assumption of the shared SED, we expected the PSF magnitudes of $r$ or $i$ bands to follow a linear function of the broad \ha{} magnitude, $-2.5\clog F_\ha{}$, with slopes of unity. Indeed, the slopes obtained from ordinary least-square fits were consistent with unity within 1$\sigma$ uncertainty in both bands, as presented in Figure~\ref{fig:sed}. Fixing the slope to unity, the best-fit relations read
\begin{eqnarray}
% \begin{aligned}
r_\mathrm{PSF} = -2.5\clog \frac{F_{\ha{}}}{10^{-6.25\pm0.01}\,\mathrm{erg\,s^{-1}\,cm^{-2}}}\label{eq:rmag}
\\
i_\mathrm{PSF} = -2.5\clog \frac{F_{\ha{}}}{10^{-6.30\pm0.01}\,\mathrm{erg\,s^{-1}\,cm^{-2}}}\label{eq:imag} 
% \end{aligned}
\end{eqnarray}
We note that the relation did not change significantly even when we included all 7826 objects from the quasar subset.

We used the SED to calculate the saturation limit of broad \ha{} flux. First, we found that for the subset of pure AGNs \begin{eqnarray}
% \begin{aligned}
r_\mathrm{fiber}-r_\mathrm{PSF} = 0.31\pm0.01\,[\mathrm{mag}] 
\\
i_\mathrm{fiber}-i_\mathrm{PSF} = 0.30\pm0.01\,[\mathrm{mag}] 
% \end{aligned}
\end{eqnarray}
Then, saturation limits in both bands can be expressed as $i_\mathrm{PSF}<14.2$ or $r_\mathrm{PSF}<14.7$. Converting both magnitudes using Eqs.~\ref{eq:rmag} and \ref{eq:imag}, we derived the saturation limit of \ha{} to be the smaller of two values, $7.5\times10^{-13}\,\mathrm{erg\,s^{-1}\,cm^{-2}}$. We removed any AGN in our sample above this threshold as described in Section~\ref{s:method}.

\subsection{The Flux Selection Function}\label{ss:selfunc}
While a majority of our sample is selected from the SDSS quasar sample, a subsample of 4308 objects are not in the quasar sample because the host galaxy flux dominates over AGN continuum in the broadband photometry. However, they are distinctively AGNs as demonstrated by a prominent broad \ha{} emission line. Since the quasar sample excluded any extended object whose color was similar to that of galaxies \citep{Richards+2002}, these objects were included in the galaxy sample because they appeared as galaxies in $r$-band images \citep{Strauss+2002}. Therefore, to assign a detection probability for these AGNs based on the AGN property, it is necessary to adopt a correlation between AGN and host galaxy properties. Specifically, for a given flux of broad \ha{}, we need to constrain the probability of being included in the galaxy sample by having $r_\mathrm{Petro}\leq17.7$, which we express as $P\condp{r_\mathrm{Petro}\leq 17.7}{F_\mathrm{b\ha}}$.

\begin{figure*}[ht!]
\centering
\includegraphics[width=0.9\textwidth]{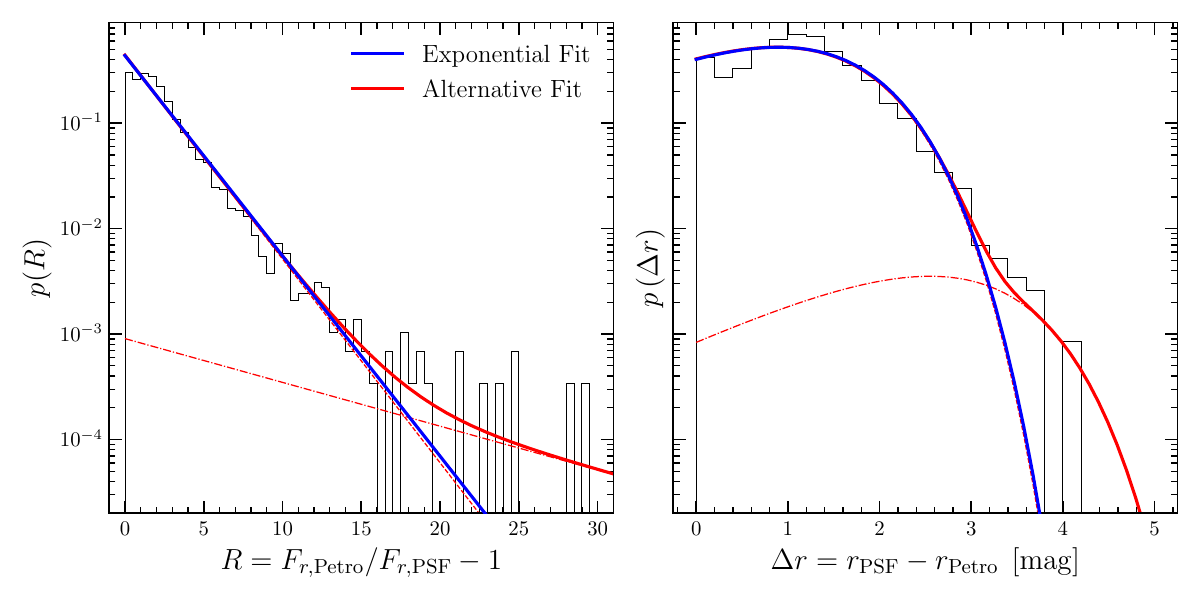}
\caption{Probability distribution of the flux ratio between Petrosian and PSF magnitudes (left) and the magnitude differences $\Delta r \equiv r_\mathrm{PSF} - r_\mathrm{Petro}$ (right). The normalized histograms are denoted with black bars. The best fit for an exponential distribution (Eq.~\ref{eq:standardselfun}) is shown with a blue solid line, while the alternative fit (Eq.~\ref{eq:altsel}) is expressed with a red solid line, along with its two exponential components (dashed and dotted lines).
\label{fig:r_petro_vs_psf}}
\end{figure*}

We first compared the $r_\mathrm{Petro}$ against the $r_\mathrm{PSF}$ for our entire sample. We found that the ratio of ``extended flux'' in $r$ band to the PSF flux, $R \equiv 10^{-0.4 \left(r_\mathrm{Petro}- r_\mathrm{PSF}\right)} - 1$,  follows an exponential distribution,
\begin{eqnarray}
\begin{aligned}
p\condp{R}{k} &= \left\{\begin{matrix}
k \exp \left[-k R\right] & \qquad (R\geq0)
\\
0 & \qquad (R<0)
\end{matrix}\right.\label{eq:standardselfun}
\end{aligned}
\end{eqnarray}
as shown in Figure~\ref{fig:r_petro_vs_psf}. The best fit reads $k = 0.44\pm0.01$. This, in turn, can be used to compute the probability of observing $r_\mathrm{Petro}$ for an AGN given the value of $r_\mathrm{PSF}$:
\begin{eqnarray}
\begin{aligned}
P&\condp{r_\mathrm{Petro}\leq 17.7}{r_\mathrm{PSF}} \\&= \int_{R=10^{-0.4\left(17.7-r_\mathrm{PSF}\right)-1}}^{\infty}p\condp{R}{k}\,dR
\label{eq:galselfunc}
\end{aligned}
\end{eqnarray}

\begin{figure}[ht!]
\centering
\includegraphics[width=0.45\textwidth]{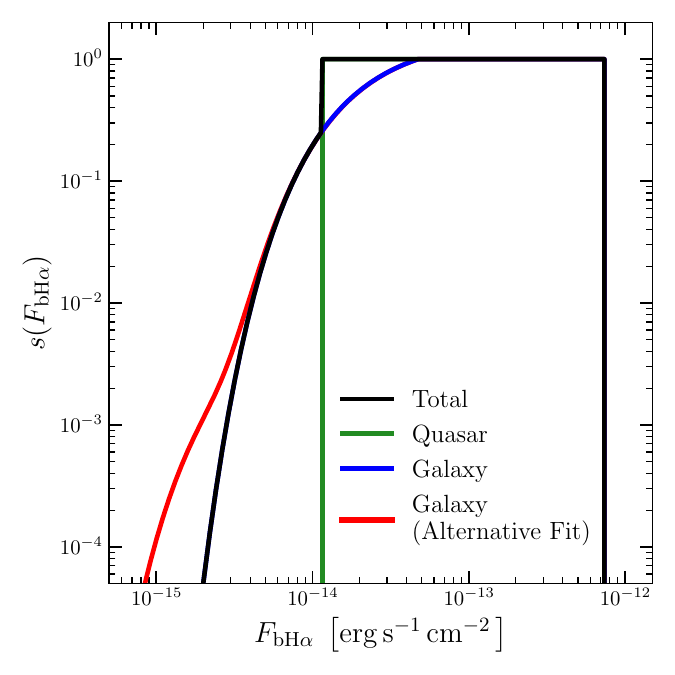}
\caption{Selection function of AGNs with a broad \ha{} line (see Section~\ref{ss:selfunc}). The selection function used for modeling the BHMF is denoted with a solid black line by combining quasar and galaxy samples, while the selection function of the quasar sample is presented with a green line. The probability that an AGN is included in the galaxy sample (Eq.~\ref{eq:galselfunc}) is expressed with a solid blue line, while the selection function constructed based on the alternative fit (Eq.~\ref{eq:altsel}) is shown by a red solid line.
\label{fig:selfunc}}
\end{figure}

Based on this, we calculated our fiducial selection function at a given broad \ha{} flux, $s(F_\mathrm{b\ha})$, as follows:
\begin{enumerate}
\item $s(F_\mathrm{b\ha})=0$ if $F_\mathrm{b\ha}>7.5\times10^{-13}\,\mathrm{erg\,s^{-1}\,cm^{-2}}$.
\item $s(F_\mathrm{b\ha})=0$ if $F_\mathrm{b\ha}<10^{-16}\,\mathrm{erg\,s^{-1}\,cm^{-2}}$.
\item Calculate the $i_\mathrm{PSF}$ using Eq.~\ref{eq:imag}.
\item $s(F_\mathrm{b\ha})=1$ if $i_\mathrm{PSF}\leq19.1$.
\item If $i_\mathrm{PSF}>19.1$, calculate $r_\mathrm{PSF}$ using Eq.~\ref{eq:rmag}.
\item $s(F_\mathrm{b\ha})=P\condp{r_\mathrm{Petro}\leq 17.7}{r_\mathrm{PSF}}$
\end{enumerate}
The resulting selection function is demonstrated in Figure~\ref{fig:selfunc}.

To test the susceptibility of our fitting result to the selection function, we obtained an alternative fit with a mixture of two exponential distributions,
\begin{eqnarray}
\begin{aligned}
&p\condp{R}{k_1,\, k_2,\, f}  \label{eq:altsel}
\\&\quad= \left\{\begin{matrix}
\frac{1}{1+f}k_1 e^{-k_1 R} + \frac{f}{1+f}k_2 e^{-k_2 R} & \qquad (R\geq0),
\\
0 & \qquad (R<0)
\end{matrix}\right.
\end{aligned}
\end{eqnarray}
which can replace $p\condp{R}{k}$ in Eq.~\ref{eq:galselfunc}. We determined the best-fit parameters in Eq.~\ref{eq:altsel} as $k_1=0.44\pm0.01$, $k_2=0.10\pm0.02$, and $f=0.01\pm0.01$, where the uncertainties were calculated based on 10000 bootstrap repetitions (see red lines in Figure~\ref{fig:r_petro_vs_psf}). Using this best fit, we also calculated the alternative selection function as shown in Figure~\ref{fig:selfunc} (magenta line). We discuss the effect of the choice of the selection function on the BHMF in Section~\ref{ss:comparison}.

%%%%%%%%%%%%%%%%%%
% Methods        %
%%%%%%%%%%%%%%%%%%

\section{Methods}\label{s:method}
In this section, we present the method for determining the BHMF and ERDF. We describe maximum posterior modeling in Section~\ref{ss:maxpost} and then we elaborate the $1/V_\mathrm{max}$ method in Section~\ref{ss:vmax}.

\subsection{Maximum Posterior Modeling}\label{ss:maxpost}
We model the distribution function of the number density of black holes at a given mass and Eddington ratio, $\phi(\mbh{},\,\Lambda)$, using a joint distribution of the black hole mass (\mbh{}) and the Eddington ratio ($\Lambda\equiv L_\mathrm{bol}/L_\mathrm{Edd}$). By marginalizing the other distribution function, the BHMF ($\phi(\mbh)$) and the ERDF ($\phi(\Lambda)$) can be respectively expressed as
\begin{eqnarray}
% \begin{aligned}
\phi(\mbh) = \int_{0}^{\infty} \phi(\mbh{},\, \Lambda) \, d\Lambda ,
\\
\phi(\Lambda) = \int_{0}^{\infty} \phi(\mbh{},\, \Lambda) \, d\mbh .
% \end{aligned}
\end{eqnarray}
While $\phi$ is a function of $(\mbh{},\, \Lambda)$, it can be described with any other pair of variables if the pair can fully describe $(\mbh{},\, \Lambda)$ (i.e., the Jacobian determinant between the pairs is not zero). For example, the LF is expressed as
\begin{eqnarray}
\begin{aligned}
\phi(L_\mathrm{bH\alpha}) = \int_{0}^{\infty} \phi\left(\mbh{},\, \Lambda=\frac{L_\mathrm{bH\alpha}}{L_\mathrm{bH\alpha, Edd}}\right) d\mbh
\end{aligned}
\end{eqnarray}
For mathematical convenience, we use log-variables $(m\equiv\clog\mbh{}/\msun{}, \lambda\equiv\clog\Lambda)$ as our coordinates, i.e., $\phi(m) = \phi(\mbh{})\cdot\left(\mbh{}\ln 10 \right)$, $ \phi(\lambda) = \phi(\Lambda)\cdot\left( \Lambda \ln 10\right)$. Similarly, the luminosity is expressed as
\begin{eqnarray}
\begin{aligned}
l=\clog \frac{L_\mathrm{bH\alpha}}{\left[\mathrm{erg\,s^{-1}}\right]} = m + \lambda + k_\mathrm{Edd}
\end{aligned}
\\
\begin{aligned}
k_\mathrm{Edd} &=\clog \left(\frac{L_\mathrm{bH\alpha}}{L_\mathrm{bol}}\cdot\frac{L_\mathrm{Edd}/\left[\mathrm{erg\,s^{-1}}\right]}{\mbh{}/\msun{}}\right)
\\&= \clog \left(\frac{1.26\times10^{38}}{190}\right)
\end{aligned}
\end{eqnarray}
where the bolometric correction $L_\mathrm{bol}/L_\mathrm{bH\alpha}=190$ is adopted from \citet{Cho+2023}.

We assume that $\phi(m, \lambda)$ can be fully described by a set of parameters $\theta$. Then, the probability of any black hole having $m$ can be described as
\begin{eqnarray}
\begin{aligned}
p\condp{m}{\theta}= \frac{\phi(m)}{\int_{-\infty}^{\infty} \phi(m)\,dm}=\frac{\phi(m)}{\phi_\bullet}\label{eq:probmass}
\end{aligned}
\end{eqnarray}
where
\begin{eqnarray}
\begin{aligned}
\phi_\bullet = \int_{-\infty}^{\infty} \phi(m)\, dm = \int_{-\infty}^{\infty} \phi(\lambda)\, d\lambda\label{eq:totaldens}
\end{aligned}
\end{eqnarray}
is the comoving number density of the black holes. The probability for $\lambda$ or $l$ can be defined in a similar manner. If we assume no redshift evolution of the number density between $z_\mathrm{min}=0$ and $z_\mathrm{max}=0.35$, the probability of a black hole having a redshift $z$ should be proportional to the comoving volume at a given redshift:
\begin{eqnarray}
\begin{aligned}
p\condp{z}{\theta} = p(z) = \frac{\frac{dV}{dz}}{\int_{z_\mathrm{min}}^{z_\mathrm{max}}\frac{dV}{dz}dz}
\end{aligned}
\end{eqnarray}
where $dV/dz = 4\pi d_c^2(z) \cdot (d/dz) d_c(z)$ is the differential comoving volume given the comoving distance $d_c(z)$.

Given a set of observations $\mathbf{q}_i = \left(l_i, m_i, z_i\right)$, the posterior $p\condp{\theta}{\mathbf{q}_i}$ is written as \citep[e.g.,][]{Kelly+2009, Schulze&Wisotzki2010}
\begin{eqnarray}
\begin{aligned}
p\condp{\theta}{\mathbf{q}_i} = p(\theta)\left[p\condp{I=1}{\theta}\right]^{-n}\times \prod_{i=1}^{n} p\condp{\mathbf{q}_i}{\theta}
\end{aligned}
\end{eqnarray}
where $p(\theta)$ is the prior distribution for $\theta$, which is discussed in Section~\ref{sss:modelfunc}. The detection probability, $p\condp{I=1}{\theta}$, is a probability of any black hole being observed, under the model given by $\theta$, thus,
\begin{eqnarray}
\begin{aligned}
&p\condp{I=1}{\theta} \\&= \int_{z_\mathrm{min}}^{z_\mathrm{max}} \int_{-\infty}^{\infty} s(\fha(l,z))\,p\condp{l}{\theta}\,p\condp{z}{\theta}\, \,dl\,dz
\end{aligned}
\end{eqnarray}
where $s(\fha)$ is the flux selection function of our sample described in Section~\ref{ss:selfunc}. The luminosity integral can be expanded in terms of the integrals of $m$ and $\lambda$,
\begin{eqnarray}
\begin{aligned}
&\int_{-\infty}^{\infty} s(\fha(l,z))\,p\condp{l}{\theta}\,dl
\\&=\int_{-\infty}^{\infty}\int_{-\infty}^{\infty} s(\fha(m+\lambda+k_\mathrm{Edd},z))\,p\condp{m,\lambda}{\theta}\,dm\,d\lambda
\end{aligned}
\end{eqnarray}
We note that $p\condp{I=1}{\theta}$, which is adopted from \citet{Kelly+2009}, is analogous to the normalization described by \citet[Eq. 5]{Schulze+2015}.

The last term in the posterior, $\prod_{i=1}^{n} p\condp{\mathbf{q}_i}{\theta}$, is the product of likelihoods for individually observed AGNs, and its calculation is described in Section~\ref{sss:likelihood}.

Lastly, we constrain the comoving number density ($\phi_\bullet$) by matching the number of AGNs in our sample ($N_\mathrm{obs}$) with the number of AGNs expected to be observed given the joint distribution function and the selection function,
\begin{eqnarray}
\begin{aligned}
N_\mathrm{obs} &=\phi_\bullet \cdot \frac{\Omega^{s}}{4\pi} \int_{-\infty}^{\infty} dm \int_{-\infty}^{\infty} d\lambda \int_{z_\mathrm{min}}^{z_\mathrm{max}} dz
\\&p(m,\lambda|\theta)\cdot s(\fha(m+\lambda+k_\mathrm{Edd},z))\, \frac{dV}{dz}
\end{aligned}\label{eq:Nobs}
\end{eqnarray}
where $\Omega^{s}$ is the survey area in units of sr. Therefore, we find
\begin{eqnarray}
\begin{aligned}
\phi_\bullet &=\frac{4\pi}{\Omega^{s}}\frac{N_\mathrm{obs}}{\int_{z_\mathrm{min}}^{z_\mathrm{max}}\frac{dV}{dz}dz}\frac{1}{p\condp{I=1}{\theta}} = \frac{N_\mathrm{obs}/V_\mathrm{survey}}{p\condp{I=1}{\theta}}
\label{eq:numdens}
\end{aligned}
\end{eqnarray}
where ${V_\mathrm{survey}}=\frac{\Omega^s}{4\pi}\int_{z_\mathrm{min}}^{z_\mathrm{max}} \frac{dV}{dz} dz$ is the survey volume. We note that \citet{Kelly+2009} modeled the total number of AGNs in the survey volume using the negative binomial distribution by assuming a logarithmic prior for the total number of AGNs ($p(\log N) \sim\mathrm{const.}$). For a fixed set of parameters ($\theta$), one can show that the expected value of this approach is identical to Eq.~\ref{eq:numdens}.

\subsubsection{Incorporating Black Hole Masses and Associated Uncertainties to Likelihood}\label{sss:likelihood}

As mentioned in Section~\ref{s:sample}, we used the single-epoch mass estimator based on the luminosity and the FWHM of the \ha{} line derived by \citet{Cho+2023}, which provides accurate mass estimates for the low-mass black holes. This is written as
\begin{eqnarray}
\begin{aligned}
&\frac{\mbh}{10^6 \msun} =3.2 \times \left(\frac{\rm FWHM_{\rm H\alpha}}{10^3\,\rm km\,s^{-1}}\right)^2\left(\frac{L_{\rm H\alpha}}{10^{42}\,\rm erg\,s^{-1}}\right)^{0.61} \label{eq:mass_fwhm}
\end{aligned}
\end{eqnarray}
The scatter for this relation $\sigma$ is given by $\sigma = \sqrt{\sigma_\mathrm{SL}^2+\sigma_{f}^2}=\sqrt{0.28^2+0.12^2}=0.30$ [dex], where $\sigma_\mathrm{SL}=0.28$ is the intrinsic scatter of the size-luminosity relation and $\sigma_f=0.12$ is the uncertainty of the $f$-factor used in the mass estimator \citep{Woo+15}. We assume that the scatter follows a normal distribution in logarithmic units. We further consider the observational uncertainty of the luminosity as follows. Let $\tilde{l}$ be the \emph{true} luminosity of an AGN. We assume that the observed luminosity follows a lognormal distribution, $l_i\sim\mathcal{N}\left(\tilde{l}, {\varepsilon_i}^2\right)$. Then, the \emph{true} mass and Eddington ratio of the black hole, $\tilde{m}$ and $\tilde{\lambda}$, can be written as
\begin{eqnarray}
\begin{aligned}
\tilde{m} = m_i + B\left(\tilde{l}-l_i\right) + t
\end{aligned}
\\
\tilde{\lambda} = \lambda_i + (1-B)\left(\tilde{l}-l_i\right)-t
\end{eqnarray}
where $B=0.61\pm0.04$ is the slope of the size-luminosity relation and $t\sim\mathcal{N}\left(0, \sigma^2\right)$ is the intrinsic scatter. If we write the Gaussian kernel of width $w$ as $K_w(x) = \frac{1}{\sqrt{2\pi}w}\exp\left[-\frac{1}{2}\left(\frac{x}{w}\right)^2\right]$, the likelihood can be written as
\begin{eqnarray}
\begin{aligned}
&p\condp{\mathbf{q}_i}{\theta} = p\condp{z}{\theta}\int_{-\infty}^{\infty} d\tilde{l} \int_{-\infty}^{\infty} dt
\\&
\quad\cdot p\condp{\tilde{m}, \tilde{\lambda}}{\theta}\cdot K_\sigma(t) K_{\varepsilon_i}\left(\tilde{l}-l_i\right)
\end{aligned}
\end{eqnarray}
In practice, the Gaussian convolutions of the likelihood were performed using 11th-order Gauss--Hermite quadratures \citep{Abramowitz_Handbook}. The detection probability was calculated with fast Fourier transform with $3\times10^{-5}$~dex resolution, using the \texttt{pyFFTW} wrapper for the \texttt{FFTW3} library \citep{FFTW3, pyFFTW}.

\subsubsection{Modeling the Distribution Functions}\label{sss:modelfunc}
Given that the maximum posterior method is a parametric approach, we must express the model for $\phi(m,\lambda)$  with a set of parameters $\theta$. For instance, \citet{Kelly+2009} used a mixture of bivariate Gaussians to model the BHMF and ERDF. However, this approach requires the number of Gaussians to be large enough to accurately model the shape of the functions, especially when the dynamic range is wide. Instead, we chose our model to follow specific functions commonly used to describe luminosity and mass functions, to reduce the number of parameters to describe them.

We assume that the BHMF is independent of the Eddington ratio and the ERDF is independent of the black hole mass, similar to many studies in the literature \citep[e.g.,][]{Schulze&Wisotzki2010, Ananna+2022a}. Then, we can write the joint distribution function ($\phi(m,\lambda)$) as
\begin{eqnarray}
\begin{aligned}
\phi(m, \lambda) = \phi_\bullet \cdot p\condp{m}{\theta} \cdot p\condp{\lambda}{\theta}
\end{aligned}
\end{eqnarray}

We use the following three functions to model the shapes of the BHMF and ERDF in this study. For consistency with the literature, we present them in terms of $X$, which can be either $\mbh$ or $\Lambda$. For probabilities for log-scale variable $x = \clog X$, the probability shape changes as
\begin{eqnarray}
\begin{aligned}
p\condp{x}{\theta} = p\condp{X}{\theta} \cdot X \ln 10
\end{aligned}
\end{eqnarray}

A \citet{Schechter1976} function (abbreviated as `S') is motivated by the galaxy LF:
\begin{eqnarray}
\begin{aligned}
p\condp{X}{\theta} \sim \frac{1}{X_c}\left(\frac{X}{X_c}\right)^\alpha \cdot \exp\left(-\frac{X}{X_c}\right)
\end{aligned}
\end{eqnarray}
This function is described by two parameters (apart from the comoving number density), $X_c$ and $\alpha$.

A modified Schechter function (abbreviated as `mS') is also used \citep[e.g.,][]{Aller&Richstone2002, Shankar+2004, Schulze&Wisotzki2010}:
\begin{eqnarray}
\begin{aligned}
p\condp{X}{\theta} \sim \frac{1}{X_c}\left(\frac{X}{X_c}\right)^\alpha \cdot \exp\left(-\left[\frac{X}{X_c}\right]^\beta\right)
\end{aligned}
\end{eqnarray}
which adds another parameter, $\beta$, to the Schechter function.

Lastly, motivated by the double power law used in the literature \citep[e.g.,][]{Schulze&Wisotzki2010}, we define a broken power law (abbreviated as `bpl') as
\begin{eqnarray}
\begin{aligned}
p\condp{X}{\theta} \sim\frac{1}{X_c}\left(\frac{X}{X_c}\right)^{\gamma_1} \left(\frac{1}{2}\left[1+\frac{X}{X_c}\right]\right)^{\gamma_2-\gamma_1}
\end{aligned}
\end{eqnarray}
This function was chosen over the double power-law function because the broken power-law function does not suffer from degeneracy between the power indices $\gamma_1$ and $\gamma_2$; $\phi\sim X^{\gamma_1}$ always holds when $X\ll X_c$ while $\phi\sim X^{\gamma_2}$ when $X\gg X_c$.

Integrating these functions over $X\in[0,\infty)$, or equivalently $x\in(-\infty,\infty)$, usually diverges, unless specific combinations of parameters are used. Thus, we normalize these functions by restricting our integration within the following intervals: $m\in[4, 11]$ and $\lambda\in[-3,0.5]$. As a result, our derived BHMF only describes the distribution of (type 1) ``active'' black holes.

We imposed flat priors for the parameters describing the models, bounded by intervals to be considered physical. For the parameters describing characteristic scales ($M_c$ or $\Lambda_c$), the priors were chosen so that they are flat in the logarithmic scale. When the maximum posterior estimate was found near the edge of the interval, we relaxed the constraining interval by several dex and performed our analysis again. If it still failed to be constrained, we simply presented the last result as is and described the convergence.

In this study, we consider three pairs of two fitting functions for the BHMF and ERDF: (1) a combination of a broken power law for the BHMF and a modified Schechter function for the ERDF (hereafter bpl--mS), which is our best-fit model; (2) a combination of a modified Schechter function for the BHMF and a Schechter function for the ERDF (hereafter mS--S), which is motivated by the work of \citet{Schulze&Wisotzki2010}; and (3) a combination of broken power laws for both the BHMF and the ERDF (hereafter bpl--bpl), which is adopted from \citet{Ananna+2022a}. We will discuss the differences in results using these models in Section~\ref{ss:comparison}.

\subsection{$1/V_\mathrm{max}$ and $1/V_\mathrm{survey}$ Methods}\label{ss:vmax}
While we mainly use maximum posterior modeling for constraining the BHMF and ERDF, we also use the traditional $1/V_\mathrm{max}$ method \citep{Schmidt1968} for comparison. Nonparametric estimates of the BHMF and ERDF can be obtained by simply counting the effective number of sources within a bin of mass, Eddington ratio, or luminosity and then dividing the counts by the survey volume $V_\mathrm{survey}$. The effective number of an object is inflated by a factor of $V_\mathrm{survey}/V_{\mathrm{max},i}$, where $V_{\mathrm{max},i}$ represents the volume within which the object in question can be observed given the selection function. However, the $1/V_\mathrm{max}$ method fails to correct for the observational uncertainties, leading to Eddington bias at high luminosities \citep{Eddington1913}. Furthermore, because the selection is based on the flux, not $\mbh$ or $\lambda$, the derived BHMF or ERDF is biased against the low-mass or low Eddington ratio objects, which is referred to as ``sample censorship'' or ``sample truncation'' in the literature \citep[e.g.,][]{Schulze&Wisotzki2010, Ananna+2022a}. Despite these limitations, this method is popularly used in the literature because of its simplicity. 

The average function, $\left\langle\phi_x\right\rangle$, within an interval $x\in[x_l, x_u]$ for an arbitrary variable $x$ is given as
\begin{eqnarray}
\begin{aligned}
\left\langle\phi_x\right\rangle &= \frac{1}{x_u-x_l}\int_{x_l}^{x_u} \phi(x)\,dx
\\&= \frac{1}{x_u-x_l}\sum_{x_i\in\left[x_l, x_u\right]}\frac{1}{V_{\mathrm{max}, i}}
\label{eq:Vmax_pointwise}
\end{aligned}
\end{eqnarray}
and its uncertainty is
\begin{eqnarray}
\begin{aligned}
\delta\left[\left\langle\phi_x\right\rangle\right] = \frac{1}{x_u-x_l}\sqrt{\sum_{x_i\in\left[x_l, x_u\right]}\frac{1}{V_{\mathrm{max}, i}^2}} 
\label{eq:Vmax_uncert}
\end{aligned}
\end{eqnarray}
We calculate $V_{\mathrm{max}, i}$ for each object as
\begin{eqnarray}
\begin{aligned}
V_{\mathrm{max}, i} = \frac{\Omega^{s}}{4\pi}\int_{z_\mathrm{min}}^{z_\mathrm{max}} p\condp{I=1}{x_i, z}\cdot\frac{dV}{dz} dz \label{eq:Vmax}
\end{aligned}
\end{eqnarray}
where $p\condp{I=1}{x_i, z}$ is the selection probability for an object having $x_i$. For instance, if $x_i=l_i$, the selection probability becomes the flux selection function $p\condp{I=1}{l_i, z} = s\left(\fha(l_i, z)\right)$, which we used for calculating the ``flux-corrected'' estimate of the LF. On the other hand, $p\condp{I=1}{x_i, z} = 1$ within comoving distances $d_c\in\left[d_{c, \mathrm{min}, i}, d_{c, \mathrm{max}, i}\right]$ reduces $V_{\mathrm{max},i}$ to the volume of a truncated cone \citep[e.g.,][]{Weigel+2016} as
\begin{eqnarray}
\begin{aligned}
&V_{\mathrm{max}, i} = \frac{\Omega^{s}}{3}\left(d_{c, \mathrm{max}, i}^3 - d_{c, \mathrm{min}, i}^3\right)
\end{aligned}
\end{eqnarray}

To correct for the sample censorship, we follow the recipe described by \citet{Schulze&Wisotzki2010}. The selection probability for an AGN at a certain black hole mass or Eddington ratio is computed as
\begin{eqnarray}
\begin{aligned}
&p\condp{I=1}{m_i, z}
\\&\quad\quad\quad\quad= \int_{\lambda_\mathrm{min}}^{\lambda_\mathrm{max}} s(\fha(l(m_i, \lambda),z))\,p(\lambda)\,d\lambda\label{eq:pIm}
\end{aligned}
\\
\begin{aligned}
&p\condp{I=1}{\lambda_i, z}
\\&\quad\quad\quad\quad= \int_{m_\mathrm{min}}^{m_\mathrm{max}} s(\fha(l(m, \lambda_i),z))\,p(m)\,dm\label{eq:pIlambda}.
\end{aligned}
\end{eqnarray}
Replacing the flux selection function in Eq.~\ref{eq:Vmax} with the mass selection probability or the Eddington ratio selection probability yields ``mass-corrected'' or ``Eddington-ratio-corrected'' estimates. Note that the mass-corrected estimate requires \emph{a priori} knowledge of the ERDF, while the Eddington-ratio-corrected estimate requires \emph{a priori} knowledge of the BHMF, in contrast to flux-corrected estimates \citep[see discussion by][]{Schulze&Wisotzki2010}. In addition, the mass-corrected estimates still suffer from the Eddington bias. Therefore, we use $1/V_\mathrm{max}$ estimates for a consistency check. 

Alternatively, we define another metric, namely, $1/V_\mathrm{survey}$, which can be obtained by replacing $V_{\mathrm{max},i}$ in Eqs.~\ref{eq:Vmax_pointwise} and \ref{eq:Vmax_uncert} with $V_\mathrm{survey}$. Effectively, this yields a histogram of the sample normalized by the survey volume and the interval size. Note that this histogram is not corrected for any of the biases we discussed previously. Since $1/V_\mathrm{survey}$ estimates do not depend on any assumptions, we compare them with the prediction from our models to assess the goodness of the fit. For this, we define the ``observed functions'' as
\begin{eqnarray}
\begin{aligned}
\hat{\phi}(m) &= \int_{-\infty}^{\infty} \phi(m')\left(\frac{V(m')}{V_\mathrm{survey}}\right) K_\sigma (m-m')\, dm'\label{eq:ucmf}
\end{aligned}
\\
\begin{aligned}
\hat{\phi}(\lambda) &= \int_{-\infty}^{\infty} \phi(\lambda') \left(\frac{V(\lambda')}{V_\mathrm{survey}}\right)K_\sigma(\lambda-\lambda')\,d\lambda'\label{eq:ucef}
\end{aligned}
\end{eqnarray}
where
\begin{eqnarray}
\begin{aligned}
&{V(x)}=\frac{\Omega^s}{4\pi}\int_{z_\mathrm{min}}^{z_\mathrm{max}} p\condp{I=1}{x, z}\frac{dV}{dz} dz\label{eq:vm}
\end{aligned}
\end{eqnarray}
The Gaussian kernel $K_\sigma$ in Eqs.~\ref{eq:ucmf} and \ref{eq:ucef} represents the scatter (uncertainty) of the single-epoch mass estimator. We adopt $\sigma=0.30\,\mathrm{[dex]}$ unless a different scatter is assumed for the particular model. A convolution with this kernel simulates the Eddington bias induced by the uncertainty of the single-epoch mass, as discussed in Section~\ref{sss:likelihood}. In the case of luminosities, the observational uncertainty, $\varepsilon_{i}$, is present in $1/V_\mathrm{survey}$ estimates, and the average luminosity uncertainty of our sample is $\sqrt{\left\langle {\varepsilon_i}^2\right\rangle}\sim0.03\,[\mathrm{dex}]$ ($<0.09\,[\mathrm{dex}]$), which is far smaller than the scatter of the single-epoch mass estimator. Thus, the observed functions are expected to closely replicate the $1/V_\mathrm{survey}$ estimates. We discuss the goodness of our models by comparing $1/V_\mathrm{max}$ and $1/V_\mathrm{survey}$ estimates in Section~\ref{ss:comparison}.

%%%%%%%%%%%%%%%%%%
% Results        %
%%%%%%%%%%%%%%%%%%

\section{Results}\label{s:results}
In this section, we present the BHMF and ERDF estimates by applying the methods presented in Section~\ref{s:method}. We performed Markov Chain Monte Carlo (MCMC) simulations using the \texttt{emcee} library \citep{Foreman-Mackey+2013} to obtain the maximum posterior solution. We carried out 10,000 iterations of the MCMC simulation with an ensemble size of twice the number of parameters. The set of these parameters did not include the comoving number density ($\phi_\bullet$), which is computed after the posterior distribution is obtained. By inspecting the trace plots of each model, we concluded that the parameters always converged before the first 5000 iterations. From the latter half of iterations, we chose the maximum posterior parameters and their central 68\% interval (equivalent to 1$\sigma$ for normal distribution) as our best-fit estimates and uncertainties. The comoving number density ($\phi_\bullet$) was computed for each set of parameters in the posterior sample using Eq.~\ref{eq:numdens}, and its best-fit values and uncertainties were selected. The results are summarized in Table~\ref{table:bestfit}. 

We also calculated $1/V_\mathrm{max}$ estimates for flux-corrected, mass-corrected, and Eddington-ratio-corrected cases, as well as $1/V_\mathrm{survey}$ estimates, as described in Section~\ref{ss:vmax}. We use the interval sizes of $0.5$~dex for estimating the BHMF and LF and $0.25$~dex for the ERDF. The minimum nonzero value for a $1/V_\mathrm{survey}$ estimate, corresponding to a singular object in a given interval, is $1.0\times10^{-9}\,\mathrm{Mpc^{-3}\,dex^{-1}}$ for the BHMF and $2.1\times10^{-9}\,\mathrm{Mpc^{-3}\,dex^{-1}}$ for the ERDF.

\renewcommand{\arraystretch}{1.25}
\begin{deluxetable*}{lBcBcccBccc}[!ht]
\tablewidth{0.95\textwidth}
\tablecolumns{11}
\tablecaption{The Best-fit Parameters of the BHMF Model\label{table:bestfit}}
\tablehead
{
    \colhead{Model}& \colhead{}&\colhead{$\phi_\bullet$}
    &\colhead{}&\multicolumn{3}{c}{{BHMF}}&\colhead{}&\multicolumn{3}{c}{{ERDF}}
    \\
    \colhead{}&\colhead{}&\colhead{$\left[10^{-4}\,\mathrm{Mpc^{-3}}\right]$}&\colhead{}&\colhead{$\clog M_c/\msun$}&\colhead{$\alpha$ or $\gamma_1$}&\colhead{$\beta$ or $\gamma_2$}&\colhead{}&\colhead{$\clog \Lambda_c$}&\colhead{$\alpha$ or $\gamma_1$}&\colhead{$\beta$ or $\gamma_2$}
    \\
    \colhead{(1)}&\colhead{}&\colhead{(2)}&\colhead{}&\colhead{(3)}&\colhead{(4)}&\colhead{(5)}&\colhead{}&\colhead{(6)}&\colhead{(7)}&\colhead{(8)}
}
\startdata
bpl--mS && $\verrud{5.32}{0.37}{0.66}$ & &\valerrud{6.75}{0.05}{0.07} & \valerrud{-0.60}{0.11}{0.08}&\valerrud{-2.56}{0.02}{0.03} & &\valerrud{-2.35}{0.31}{0.31} & \valerrud{0.34}{0.21}{0.22} &\valerrud{0.53}{0.03}{0.08}
\\
mS--S && $\verrud{11.0}{0.82}{0.28}$ & &\valerrud{1.01}{0.22}{0.01}\tablenotemark{\rm\dag} & \valerrud{-0.20}{0.01}{0.08} &\valerrud{0.164}{0.003}{0.001} & &\valerrud{-1.20}{0.01}{0.01} & \valerrud{-0.62}{0.03}{0.02} &-
\\
bpl--bpl && $\verrud{5.61}{0.50}{0.57}$ & &\valerrud{6.76}{0.06}{0.05} & \valerrud{-0.62}{0.09}{0.09} &\valerrud{-2.56}{0.02}{0.03} & &\valerrud{-0.41}{0.12}{0.11} & \valerrud{-0.34}{0.06}{0.08} &\valerrud{-10.25}{1.94}{1.55}
\\\hline
bpl--mS \\
\ \   Alt. Selection && $\verrud{5.27}{0.45}{0.74}$ & &\valerrud{6.77}{0.06}{0.06} & \valerrud{-0.62}{0.10}{0.09}&\valerrud{-2.56}{0.03}{0.02} & &\valerrud{-2.36}{0.24}{0.42} & \valerrud{0.33}{0.30}{0.18} &\valerrud{0.52}{0.04}{0.07}
\\
\ \   $2\times$ Int. Scatter && $\verrud{4.89}{0.49}{0.27}$ & &\valerrud{6.59}{0.06}{0.07} & \valerrud{-0.69}{0.09}{0.10}&\valerrud{-2.41}{0.02}{0.02} & &\valerrud{-2.05}{0.08}{0.01} & \valerrud{2.00}{0.00}{0.07}\tablenotemark{\rm\dag} &\valerrud{0.78}{0.04}{0.01}
\enddata
\tablecomments{Column (1): the name of the model as described in Section~\ref{sss:modelfunc}. Column (2): the comoving number density of the active black holes, defined by Eq.~\ref{eq:totaldens}. Columns (3)-(5): appropriate sets of parameters for the BHMF as defined in Section~\ref{sss:modelfunc}. Columns (6)-(8): appropriate sets of parameters for the ERDF. The representative values are the maximum posterior estimators, and the uncertainties are the central 68\% intervals. }
\tablenotetext{\dag}{Converged on parameter bounds.}
\end{deluxetable*}

\subsection{Comparison of BHMF and ERDF Estimates Depending on the Functional Forms}\label{ss:comparison}
\begin{figure*}[ht!]
\centering
\includegraphics[width=0.45\textwidth]{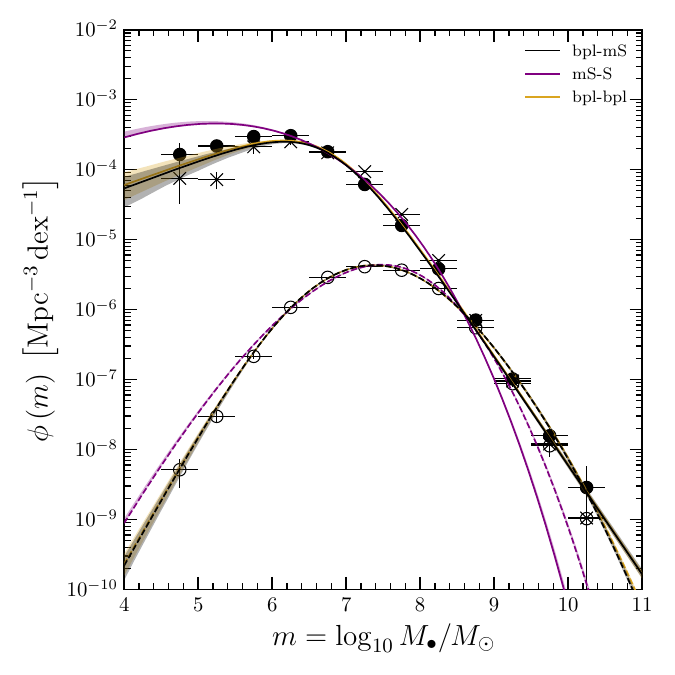}
\includegraphics[width=0.45\textwidth]{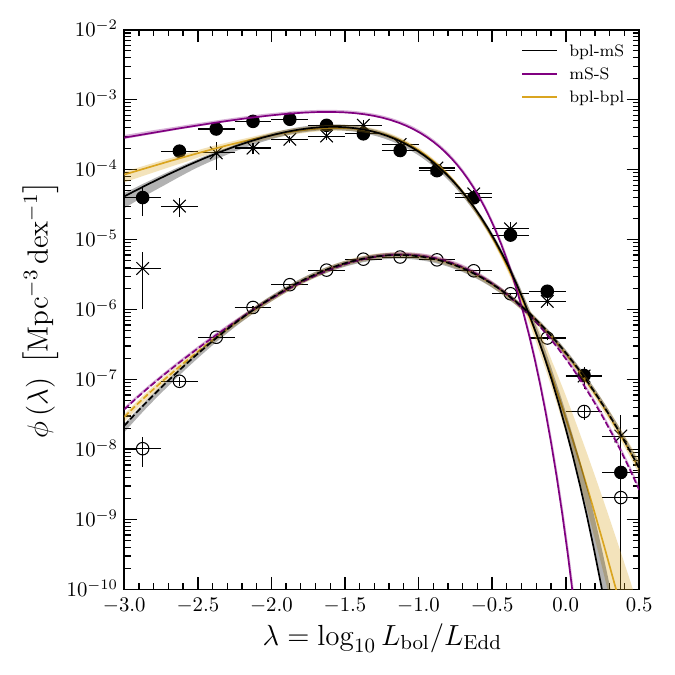}
\caption{The maximum posterior estimates and 1$\sigma$ uncertainties for the BHMF (left) and ERDF (right) based on the three different pairs of the functional forms (black, magenta, and yellow solid lines), compared with the mass-corrected (left) or Eddington-ratio-corrected (right) estimates (black filled circles) and the flux-corrected estimates (crosses) based on the $1/V_\mathrm{max}$ method with the bpl--mS model. The simulated observed density functions are presented for the BHMF (left) and ERDF (right) after considering the selection function and the mass uncertainty. The $1/V_\mathrm{survey}$ estimates (open circles) are compared with the maximum posterior simulations. 
\label{fig:models}}
\end{figure*}

We compare the maximum posterior estimates, $1/V_\mathrm{max}$ estimates, and $1/V_\mathrm{survey}$ estimates in Figure~\ref{fig:models}. The maximum posterior estimates for the intrinsic BHMF and ERDF based on the bpl--mS and bpl--bpl models are generally similar, while the mS--S model-based estimates show a different shape. This is because the exponential nature of the mS--S model cannot reproduce the shape of the observed BHMF at the high-mass end. Note that the uncertainties of $\phi(m)$ and $\phi(\lambda)$ are relatively small except for the high Eddington ratio end because of a small number in the bin. For example, there is only one object at $\lambda>0.25$. 

The $1/V_\mathrm{max}$ estimates are larger than the maximum posterior estimates, particularly for the ERDF. For example, the $1/V_\mathrm{max}$ estimate of $\phi(\lambda)$ is larger than the maximum posterior $\phi(\lambda)$ by more than 1~dex at $\lambda>0$. This is because the correction factors for the sample censorship (i.e., Eqs.~\ref{eq:pIm} and \ref{eq:pIlambda}) do not correct for the Eddington bias, while the maximum posterior estimates suffer no Eddington bias. Thus, the $1/V_\mathrm{max}$ estimates are suppressed near the maxima of $1/V_\mathrm{survey}$ estimates and enhanced elsewhere. This demonstrates the bias inherent to the $1/V_\mathrm{max}$ and $1/V_\mathrm{survey}$ estimators. Hereafter we only present $1/V_\mathrm{survey}$ estimates when comparing the observed functions without further discussing $1/V_\mathrm{max}$ estimates.

We also present the simulated observed functions for the BHMF and ERDF based on the maximum posterior estimates, after taking into account the selection function and the uncertainties of mass and Eddington ratio (dotted lines in Figure~\ref{fig:models}). Then, we compare the simulated observed functions with the $1/V_\mathrm{survey}$ estimates. As described in Section~\ref{ss:vmax}, the simulated observed functions are expected to properly reproduce $1/V_\mathrm{survey}$ estimates. We find that the simulated observed function of the BHMF based on bpl--mS and bpl--bpl models is virtually consistent with the $1/V_\mathrm{survey}$ estimates. In contrast, the simulated observed function based on the mS--S model predicts a smaller number of high-mass AGNs and a larger number of low-mass AGNs than the other two model-based simulations, as well as the $1/V_\mathrm{survey}$ estimates. Given that we aim to constrain the BHMF at the low-mass end, we conclude that the mS--S model does not properly fit the sample. 
In the case of the ERDF, the simulated observed functions based on all three models are similar and generally consistent with the $1/V_\mathrm{survey}$ estimates. At the lowest $\lambda$, the discrepancy is pronounced, with the bpl--mS model predicting the lowest number of AGNs. Considering that the $1/V_\mathrm{survey}$ estimate is even smaller than the simulation of the other two model-based estimates, we conclude that the bpl--mS model reproduces the ERDF the best, although the difference among three models is not significant. For the rest of the paper, we present the bpl--mS model-based maximum posterior estimates as the representative model.

\begin{figure*}[ht!]
\centering
\includegraphics[width=0.45\textwidth]{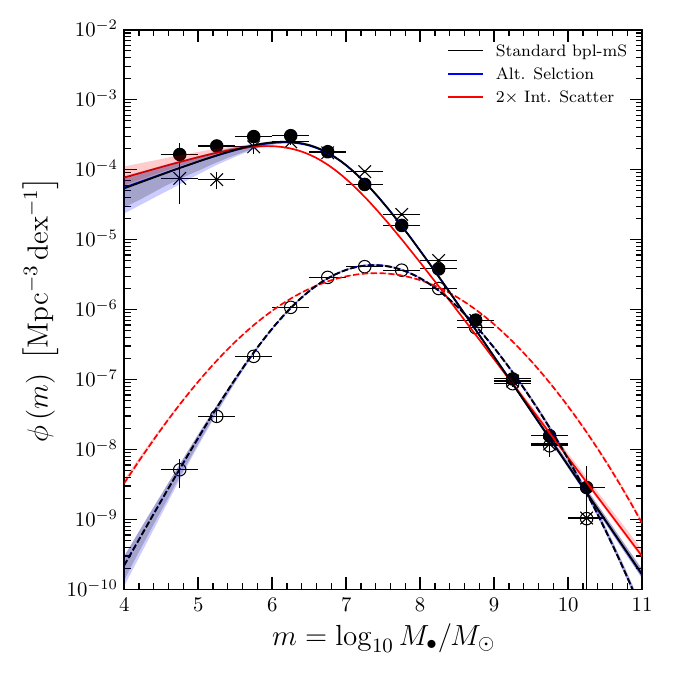}
\includegraphics[width=0.45\textwidth]{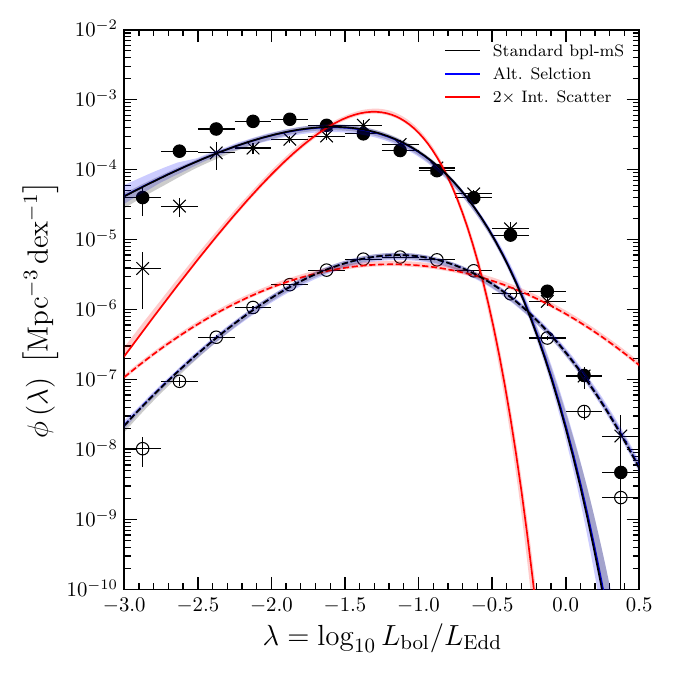}
\caption{Comparison of the maximum posterior estimates and 1$\sigma$ uncertainties for the BHMF (left) and ERDF (right) based on different assumptions and the mass uncertainty. Our best estimates of the intrinsic BHMF and ERDF based on the bpl--mS model and 0.3 dex mass uncertainty (black lines) are compared with the estimates using the alternative selection function (blue lines; described in Section~\ref{ss:selfunc}) or using a factor of two increased mass uncertainty (red lines). The simulated observed functions based on maximum posterior modeling with each of three assumptions (dotted lines) are compared with the $1/V_\mathrm{survey}$ estimates (open circles). 
\label{fig:assumptions}}
\end{figure*}

\subsection{Dependency of the Shape of the Selection Function and Mass Uncertainty}
We investigate the effect of the different assumptions that we applied to model the selection function and the uncertainty of black hole mass on the estimates of the BHMF and ERDF. We assume an uncertainty of 0.3~dex for the single-epoch mass estimator and utilize the selection function given by Eq.~\ref{eq:standardselfun} for maximum posterior modeling and the $1/V_\mathrm{max}$. We denote the maximum posterior estimates based on the bpl--mS model and the 0.3~dex mass uncertainty as ``Standard bpl--mS''. In addition, we consider two alternative cases. First, we adopt the alternative selection function described in Section~\ref{ss:selfunc}, which effectively increases the selection probability at low flux (see the red line in Figure~\ref{fig:selfunc}), and calculate the BHMF and ERDF estimates based on maximum posterior modeling. Second, we artificially increased the mass uncertainty, $\sigma$, from 0.3 to 0.6~dex (see Section~\ref{sss:likelihood}) to consider unaccounted-for uncertainties in the mass estimate, and we carried out maximum posterior modeling, which is expressed as ``$2\times$ Int. Scatter.'' In other words, we used $\sigma=0.6\,\mathrm{dex}$ for Eqs.~\ref{eq:ucmf} and \ref{eq:ucef}.

We compare the results based on the assumptions in Figure~\ref{fig:assumptions}. Note that the choice of the selection function does not change the BHMF and ERDF significantly since the increase of the selection probability of low-flux AGNs is insignificant for changing the number density even at low-mass and low Eddington ratio bins. In contrast, the increased intrinsic scatter substantially modifies the shape of the ERDF, while its effect on the BHMF is not clearly detected. Thus, we expect that the mass uncertainty is not likely to be as large as 0.6~dex.

Compared with the observed function constrained by $1/V_\mathrm{survey}$ estimates, the simulated observed function of the BHMF and ERDF from maximum posterior modeling with an assumption of 0.6 dex mass uncertainty shows a very different shape for both the BHMF and the ERDF, implying that the mass uncertainty is much smaller than 0.6 dex. Nevertheless, the change in the BHMF $\phi(m)$ caused by different assumptions is smaller than 0.5~dex over all mass scales. Thus, we conclude that our best fit for the BHMF is stable against the choice of different assumptions.

\subsection{Comparison with the Literature}\label{ss:lit}
To compare our results with the literature, we compiled the reported density functions from \citet[BHMF, LF(b\ha{}) ($1/V_\mathrm{max}$)]{Greene&Ho2007}, \citet[BHMF(all), LF(bol)]{Shankar+2009}, \citet[BHMF, ERDF (mS, S)]{Schulze&Wisotzki2010}, \citet[BHMF, ERDF]{Kelly&Shen2013}, and \citet[BHMF, ERDF, LF(X-ray) (type 1)]{Ananna+2022a}. In the case of the LF, we collected not only the broad \ha{} LF but also LFs for bolometric luminosity \citep{Shankar+2009} and X-ray \citep{Ananna+2022a}. We adopted an X-ray bolometric correction of $L_\mathrm{bol}/L_\mathrm{14\mbox{-}195\,keV}=7.4$ \citep{Ananna+2022a} and an \ha{} bolometric correction of $L_\mathrm{bol}/L_\mathrm{bH\alpha}=190$ \citep{Cho+2023} to convert the given LF to an \ha{} LF. We also note that the BHMF by \citet{Shankar+2009} is based on the galaxy LF and the black hole mass scaling relations. Thus, this is a \emph{total} BHMF rather than a type 1 active BHMF, which we present along with the rest of the literature. Uncertainties for the functions were collected whenever available. When they were provided as parameter uncertainties, we determined the 1$\sigma$ uncertainties based on the 1000 Monte Carlo realizations. Lastly, we multiplied the appropriate factor to compensate for the differences in the Hubble constant from the one we used ($h=0.72$).

\begin{figure*}[ht!]
\centering
\includegraphics[width=0.3\textwidth]{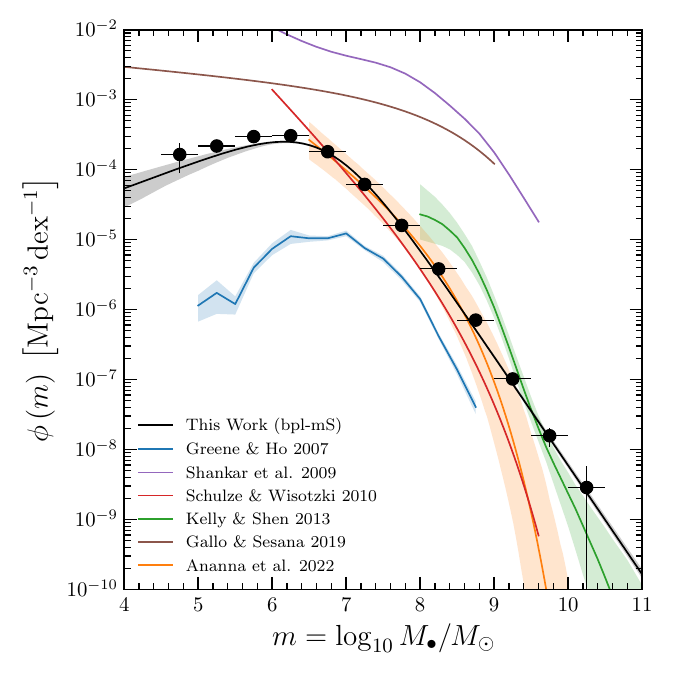}
\includegraphics[width=0.3\textwidth]{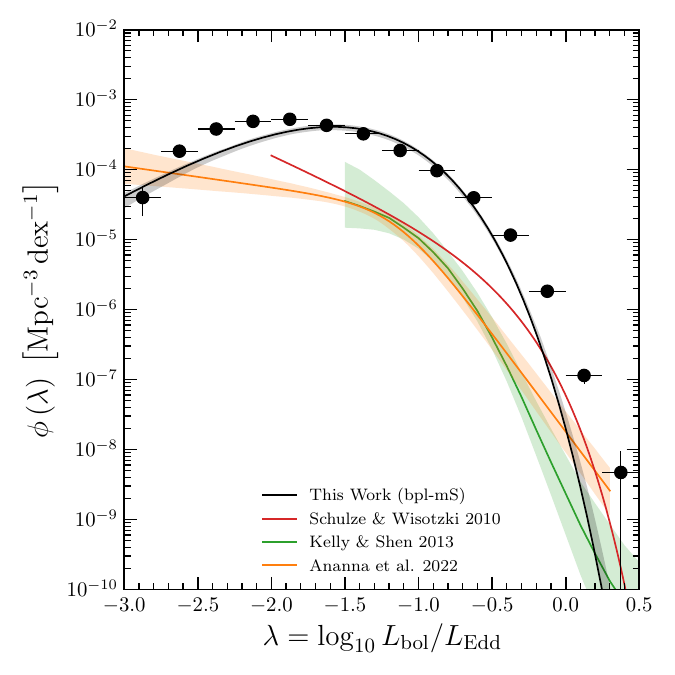}
\includegraphics[width=0.3\textwidth]{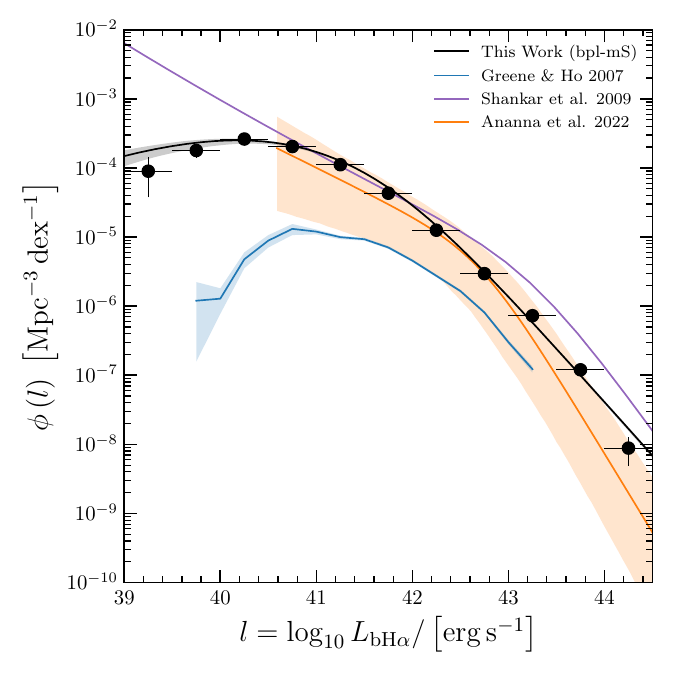}
\caption{The best fit of the BHMF (left), ERDF (middle), and broad \ha{} LF (right), compared with the literature. The maximum posterior prediction of the bpl--mS model and its 1$\sigma$ uncertainties are denoted by the black solid line and the shaded region. Filled circles denote the mass-corrected estimates obtained using the $V_{\mathrm{max}}$ method, as described in Section~\ref{ss:vmax}. Details of the individual functions from the literature are described in Section~\ref{ss:lit}.
\label{fig:bestfit}}
\end{figure*}

In Figure~\ref{fig:bestfit}, we present our best-fit BHMF, ERDF, and broad \ha{} LF along with the results collected from the literature. Regardless of the method they used, we plotted the literature functions as solid lines. We showed the functions in the intervals in which they were originally presented. If the functions were presented as values, we simply showed them all. Conversely, if they were provided as functions and parameters, we considered those functions valid only within the $x$-axis intervals of the figures plotting the functions in their respective references.

The BHMF we found follows a broken power law with a peak near the mass $\mbh\sim10^6\msun$. For the intermediate-mass regime, we find the BHMF to be relatively constant in logarithmic scale near $\sim 10^{-4}\,\mathrm{Mpc^{-3}\,dex^{-1}}$. Compared with the literature, we found a striking consistency with the functions reported by \citet{Schulze&Wisotzki2010} and \citet{Ananna+2022a} for masses $10^{6.5}\msun<\mbh<10^{9}\msun$; for higher mass, they predicted much smaller values of the BHMF, which may be because the Schechter-like models they used in their fits are asymptotic to exponential decay at the high-mass end. \citet{Kelly&Shen2013} predicted a BHMF with a similar value to our fit at higher masses ($>10^{8}\msun$) with a steeper slope, possibly due to their choice of priors on the Gaussian mixture modeling. On the contrary, we predict a significantly larger number of black holes than what \citet{Greene&Ho2007} found using the broad \ha{} from SDSS DR4 spectra. While there could be many explanations involving the spectroscopic completeness of DR4 itself, the choice of the decomposition method used by the authors, etc., the most likely explanation would be that the completeness of the sample was overestimated. This is further supported by the observation that the other studies using similar sample and selection criteria, notably \citet{Kelly&Shen2013}, who used SDSS DR7, predicted a much larger BHMF than what \citet{Greene&Ho2007} found, similar to ours. Lastly, since we constructed our BHMF based on the AGNs, our BHMF is $\sim$0.1\%--10\% of the total BHMF estimated by \citet{Shankar+2009} or \citet{Gallo&Sesana19}. We discuss the active fraction in detail in Section~\ref{ss:af}.

Our ERDF is less consistent with the literature, although this is to be expected. Since our BHMF spans a larger dynamic range than that in the literature, the total number density involved in our sample should be much higher. Consequently, the ERDF should also be much higher in proportion to the total number density. Focusing on the shape, our ERDF is more concentrated toward an Eddington ratio of a few percent, while the literature functions predict a monotonically decreasing ERDF. As demonstrated in Section~\ref{ss:comparison}, assumptions on the mass and/or luminosity uncertainty affect the resulting ERDF significantly. Nevertheless, should the ERDF be smaller for $\Lambda<$0.1-1\% as with our best-fit model, the physical mechanism behind it could be the change in the accretion state \citep[e.g.,][]{Inayoshi+2020b}. Another possible explanation could be that there is a minimum required accretion rate for a broad-line region (BLR) to exist \citep[e.g.,][]{Nicastro2000}. Then, the discrepancy in the shape between our ERDF and the X-ray-based ERDF by \citet{Ananna+2022a} can be explained as the BLR being absent for the low-$\lambda$ AGNs.

Lastly, our broad \ha{} LF is consistent with the X-ray and bolometric LF by \citet{Shankar+2009} and \citet{Ananna+2022a}. We observe the difference between our LF and the \ha{} LF by \citet{Greene&Ho2007}, and this is most likely to be of the same origin as the differences in the BHMF.

%%%%%%%%%%%%%%%%%%
% Discussions    %
%%%%%%%%%%%%%%%%%%

\section{Discussions}
\subsection{Active Fraction}\label{ss:af}
The black hole occupation fraction of low-mass galaxies provides valuable information on the origin of SMBHs \citep[e.g.,][]{Greene+20}. However, it is challenging to observationally constrain the occupation fraction because of the lack of a complete sample of IMBHs in dwarf galaxies. The active fraction is defined as a fraction of galaxies hosting actively mass-accreting black holes, providing a lower limit of the occupation fraction. Compared to the occupation fraction, the active fraction is easily obtained through AGN surveys. We measure the BHMF of AGNs across a wide dynamic range, including the intermediate-mass regime ($\mbh<10^{6}\msun$), with an overall precision of much less than 1~dex. Thus, our sample is ideal for constraining the active fraction at the low-mass end. In this section, we present the active fractions of galaxies in the local volume.

First, we investigate the correlation between the black hole mass and the total host galaxy mass. A number of previous studies presented the black hole mass correlation with bulge properties, focusing on relatively massive galaxies \citep[e.g.,][]{Kormendy&Ho13, Woo+2013}. However, the bulge properties cannot trace the total stellar mass reliably, due to the diverse range of the bulge-to-total stellar mass ratio \citep[e.g.,][]{Khochfar+2011}.

For low-mass galaxies, \citeauthor{Reines&Volonteri2015}~(\citeyear{Reines&Volonteri2015}, see also \citealt{Reines&Volonteri2015Errata}) reported a broad correlation between the black hole mass and the total stellar mass based on 271 local AGNs, including a number of IMBHs. In their study, the total stellar mass was estimated based on the color-dependent mass-to-light ratio from \citet{Zibetti+2009}. We use this sample for our analysis after updating single-epoch black hole masses based on the new estimator Eq.~\ref{eq:mass_fwhm}). Note that the previous single-epoch mass of IMBHs could be overestimated by 0.2--0.5~dex \citep[see the discussion by][]{Cho+2023}. Thus, we redetermine the black hole mass of 255 AGNs in their sample. In addition, we adopt the reverberation mass of NGC~4395 from \citet{Cho+21}. 

\begin{figure}[t!]
\centering
\includegraphics[width=0.45\textwidth]{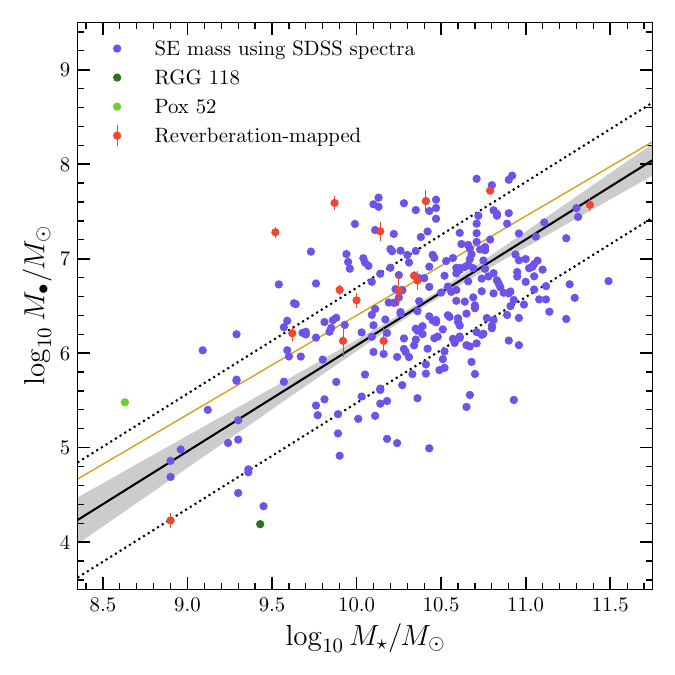}
\caption{The $\mbh$--$\mstar{}_\mathrm{,total}$ relation. The solid black line is our best fit, while the solid yellow line represents the relation presented originally by \citet{Reines&Volonteri2015}. The shaded region shows the 1$\sigma$ uncertainty of the best-fit relation, and the dotted lines mark the rms scatter around the relation (0.61~dex).
\label{fig:mmfit}}
\end{figure}

Using this sample, we obtain the $\mbh$--$\mstar{}_\mathrm{,total}$ relation as
\begin{eqnarray}
\begin{aligned}
&\clog \frac{\mbh}{\msun}
\\&= \left(7.20\pm0.08\right)+\left(1.11\pm0.12\right)\clog \frac{\mstar_\mathrm{,total}}{10^{11}\msun}
\label{eq:mmfit}
\end{aligned}
\end{eqnarray}
with an rms scatter of 0.61~dex (see Figure~\ref{fig:mmfit}).

Second, we estimate the active fraction as follows. We translate the BHMF to the corresponding mass function as a function of stellar mass using Eq.~\ref{eq:mmfit} as
\begin{eqnarray}
\begin{aligned}
\phi(M_\ast)\, d\clog M_\ast = \phi(\mbh)\, d\clog \mbh .
\end{aligned}
\end{eqnarray}
Then, the translated MF is convolved with a Gaussian kernel with a width of 0.61~dex, which is the rms scatter of the $\mbh$--$\mstar{}_\mathrm{,total}$ relation, to obtain the mass function of active galaxies. Dividing this by a galaxy mass function yields the active fraction. We adopt the galaxy mass function from \citet{Weigel+2016}.

In addition, we consider two factors that may affect the active fraction. First, we assume a type 1 fraction ($f_\text{type1}$), which is the fraction of type 1 AGNs among all AGNs. Since we use type 1 AGNs to derive the BHMF, the active fraction based on this BHMF only counts type 1 AGNs. Thus, we overcome this limitation by dividing the active fraction of type 1 AGNs by the type 1 fraction, obtaining the total active fraction. Previous studies suggest that the type 1 fraction increases with increasing black hole mass. For instance, \citet{Lu+2010} reported that the type 1 fraction is $\sim$20\% for $\mbh<10^8\msun$ and $\sim$30\% for $\mbh>10^8\msun$. Similarly, \citet{Oh+2015} found that the type 1 fraction increases from $\sim$20\% at $\mbh\sim10^6\msun$ to $\sim$60\% at $\mbh\sim10^9\msun$. In contrast, \citet{Moran+2014} found only 2 type 1 AGNs out of 28 AGNs in dwarf galaxies, corresponding to $f_\text{type1}\sim$7\%. Since we expect the type 1 fraction to be a function of the host galaxy mass or black hole mass, we assume that the upper and lower limits of the type 1 fraction are 20\% and 7\%, respectively. 

\begin{figure}[t]
\centering
\includegraphics[width=0.47\textwidth]{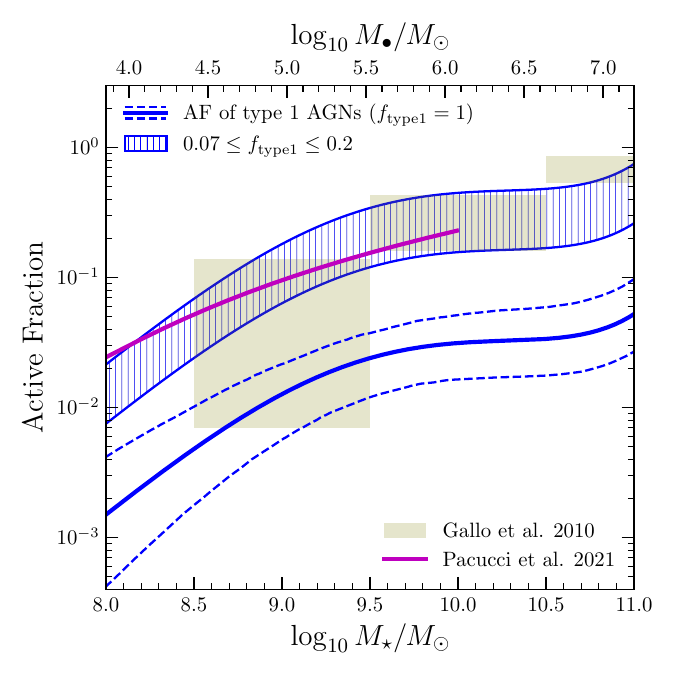}
\caption{The active fraction of galaxies based on our BHMF. The $x$-axis is the total stellar mass \mstar{} or corresponding black hole mass \mbh{} based on Eq.~\ref{eq:mmfit}. The solid blue line is based on the type 1 BHMF, with its 1$\sigma$ uncertainty denoted with dashed lines. The total active fraction, assuming type 1 fractions between 0.07 and 0.2, is shown as a hatched region. For comparison, the active fractions from \citet[beige shading]{Gallo+2010} and \citet[magenta line]{Pacucci+2021} are presented.
\label{fig:af}}
\end{figure}

We find that the active fraction of type 1 AGNs is larger than $\sim$3\% for massive galaxies with $\mstar>10^{10}\msun$ as shown in Figure~\ref{fig:af}. We also present the active fraction of all AGNs (blue shaded region), using the upper and lower limits of the type 1 fraction (i.e., 20\% and 7\%, respectively), constraining the active fraction of all AGNs to be between 40\% and 15\% for massive galaxies. In contrast, the active fraction decreases for lower-mass galaxies, particularly at $\mstar<10^{9.5}\msun$. At the low end of stellar mass ($\mstar=10^8\msun$), the active fraction is $\sim$2\%, even with a lower limit of the type 1 fraction of 7\%.

Our results are broadly consistent with the previous study based on deep X-ray observations by \citet{Gallo+2010}, who reported the active fraction to be 24\%--34\% on average, varying from 0.7\% to 87\% depending on the host galaxy mass. Note that they used very low Eddington ratio AGNs ($\lambda<-3$) for deriving the active fraction, while the lower limit of the Eddington ratio is $\lambda\geq-3$ in our sample. \citet{Gallo+2010} described that the decreasing trend of the active fraction with decreasing host galaxy mass was caused by the fact that the fixed X-ray luminosity limit of their survey corresponds to a higher Eddington ratio for lower-mass black holes in lower-mass galaxies. Nevertheless, their average active fraction of 24\%--34\% is a rough approximation for the active fraction. If we apply the lower and upper limits of the type 1 fraction (7\% and 20\%) to their active fraction, then we obtain the active fraction of type 1 AGNs as 1.68\%--6.8\%. A more recent study by \citet{Ananna+2022a} based on the X-ray luminosity reported the active fraction of 10-16\% for AGNs with $\lambda>-3$ AGNs with $10^{6.5}\leq\mbh/\msun\leq10^{10.5}$. In the case of type 1 AGNs with $\lambda>-3$, they found the active fraction to be $\sim$4\% \citep[see Figure~13 in][]{Ananna+2022a}. Thus, our result of $\sim$3\% active fraction based on type 1 AGNs is consistent with both studies.

\citet{Pacucci+2021} constrained the active fraction of galaxies with $\mstar<10^{10}\msun$, using AGNs with $\lambda>-4$, by adopting simple assumptions on the accretion efficiency. They predicted 2\%--20\% of active fraction, with an increasing trend with host galaxy mass, which is broadly consistent with our result. However, the increasing slope of their active fraction is shallower than ours but consistent with the range of active fraction that we constrained with the upper and lower limits of the type 1 fraction. Their shallower slope can be explained with a varying type 1 fraction between dwarf galaxies \citep{Moran+2014} and the more massive galaxies \citep{Lu+2010, Oh+2015}. If the BLR is less common in active IMBHs, then the type 1 fraction can be smaller for IMBHs. 

\subsection{Black Hole Occupation Fraction} \label{ss:occ}
\begin{figure}[t]
\centering
\includegraphics[width=0.47\textwidth]{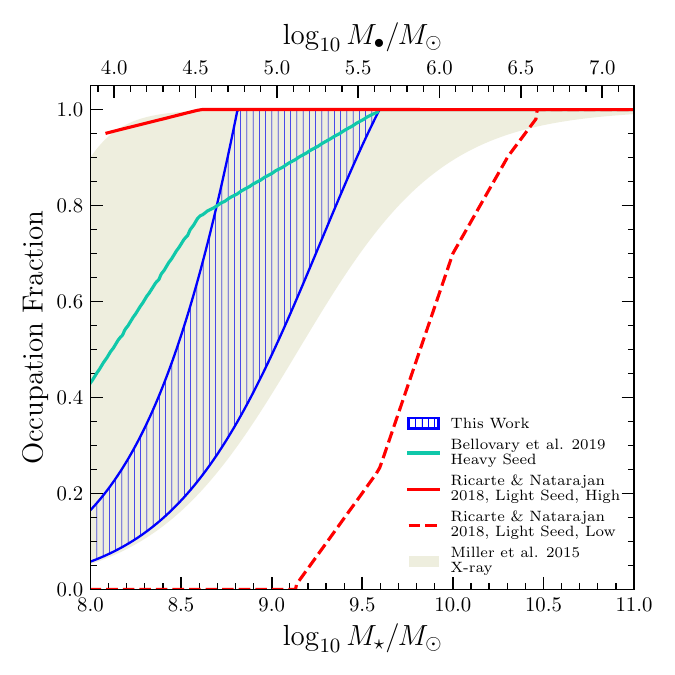}
\caption{The black hole occupation fraction of galaxies, obtained by assuming type 1 fractions between 7\% and 20\% and a duty cycle of 13\%, is shown as a hatched region. The $x$-axis is identical to that in Figure~\ref{fig:af}. The occupation fractions from the literature, as discussed in Section~\ref{ss:occ} and Section~\ref{ss:seed}, are presented by colored lines and a shaded region.
\label{fig:occ}}
\end{figure}

We derive the black hole occupation fraction from the active fraction by assuming a duty cycle, which is the fraction of active black holes among all black holes. We adopt a constant duty cycle of 13\%, which is an average of 10\% and 16\% constrained based on a sample of AGNs selected with X-ray by \citep{Ananna+2022a}. As done for the active fraction, we also use two different fractions, 7\% \citep{Moran+2014} and 20\% for type 1 AGNs\citep{Lu+2010, Oh+2015}.

We present the occupation fraction in comparison with the previous results in the literature in Figure~\ref{fig:occ}. The occupation fraction is close to 1 for galaxies with $\mstar\gtrsim10^{9.5}\msun$. This is consistent with the expectation that all massive galaxies host SMBHs \citep[e.g.,][]{Rees1984, Kormendy&Richstone1995}. The occupation fraction departs from unity for galaxies with $\mstar\lesssim10^{9.5}\msun$, becomes 0.5 for galaxies with $\mstar\sim10^{8.5}$--$10^9$\msun, and reduces to $\sim$0.1 for dwarf galaxies with $\mstar\sim10^8\msun$. Our result is broadly consistent with the occupation fraction reported by \citet{Miller+15} based on the X-ray detection, supporting the validity of the assumptions involved in the calculation of the occupation fraction.

\subsection{Comparison with the Seed Scenarios}\label{ss:seed}
We compare the occupation fraction with the theoretical predictions based on heavy- and light-seed models in the literature in Figure~\ref{fig:occ}. For the heavy-seed scenario, we adopt the occupation fraction by \citet{Bellovary+2019}, which was calculated based on a cosmological zoom-in simulation. For the light-seed scenario, we use a couple of occupation fractions by \citet[see Figure~5 of \citealt{Greene+20}]{Ricarte&Natarajan2018b}, which were calculated with two different assumptions on the accretion rate.

The occupation fractions of dwarf galaxies ($\mstar<10^{9}\msun$), predicted from the heavy-seed model \citep{Bellovary+2019} or from the optimistic light-seed model \citep[MS]{Ricarte&Natarajan2018b}, are larger than the occupation fraction we found. On the contrary, the pessimistic light-seed model \citep[PL]{Ricarte&Natarajan2018b} is far smaller than our occupation fraction.

Given the large uncertainty of the occupation fraction, it is difficult to discriminate between different seed scenarios. The uncertainty arises largely from the assumptions on the activity of black holes, i.e., the duty cycle and the type 1 fraction, that we used to construct the occupation fraction. Therefore, it is necessary to quantify the activity of IMBHs in detail, to better constrain the occupation fraction. 

\subsection{Implications for Future Surveys}
The BHMF derived in Section~\ref{s:results} can be used for estimating the number of AGNs expected in a specific survey, based on Eq.~\ref{eq:Nobs}, along with the sensitivity function of the survey. In this section, we discuss the number of IMBHs ($10^4\msun<\mbh<10^6\msun$) expected from future surveys.

For instance, the Vera C. Rubin Observatory Legacy Survey of Space and Time \citep[LSST, ][]{LSST2019} covers an area of 18,000 ${\deg}^2$ in the sky. Its limiting magnitude in $r$ band is 24.7 in a single visit and 27.5 in the final co-added survey, while objects with $r<16$ would saturate. Converting these magnitudes with Eq.~\ref{eq:rmag} and constructing a simple top-hat-like selection function, we expect that $\sim$500,000 active IMBHs within $z<0.35$ would be detected in a single visit and $\sim$2~million active IMBHs in the final survey.

One of the promising future all-sky surveys is SPHEREx \citep{SPHEREx2014}, which will cover all sky with a depth of 18.5~AB mag and a 100~${\deg}^2$ area with a depth of $\sim$22~AB-mag over the wavelength range of 0.75-5$\mu$m \citep{SPHEREx2018}. While \ha{} emissions of AGNs at $z\geq 0.143$ can be detected in the survey, AGNs at lower redshift can be detected via alternative signatures, such as Pa$\alpha$. Considering that the flux of the Pa$\alpha$ line is dimmer than that of \ha{} by a factor of 10 in type 1 AGNs \citep{Netzer90}, we can calculate the Pa$\alpha$ flux for given $\lambda$. On the other hand, AB magnitude can be derived by averaging the Pa$\alpha$ flux over the spectral resolving power $R=41$ of SPHEREx at low wavelengths. Comparing these, we derive the 5$\sigma$ limiting flux of Pa$\alpha$ to be $5.6\times10^{-14}\,\mathrm{erg\,s^{-1}\,cm^{-2}}$ for all sky and $2.2\times10^{-15}\,\mathrm{erg\,s^{-1}\,cm^{-2}}$ for a 100~deg$^2$ area. From this, we expect the number of active IMBHs detected with Pa$\alpha$ within a $z<0.143$ volume to be $\sim$25 for a 100~deg$^2$ area and $\sim$100 for all sky, without considering Galactic obstruction.

\subsection{Summary}
In this paper, we derived the mass function and ERDF of type 1 AGNs with masses down to  $10^4\msun$, using the local AGNs in SDSS DR7. By constructing the selection function based on the detection probability of the broad \ha{} line fluxes and applying maximum posterior modeling, we obtained the intrinsic and simulated density functions. We summarize the main results as follows:

\begin{enumerate}
\item The BHMF peaks at $\mbh\simeq10^6\msun$ while it is flat or slightly decreasing for IMBHs, as similarly reported in the literature.
\item The ERDF peaks near 1\%--10\% and follows a Schechter-like or a broken power-law function, exponentially decreasing as the Eddington ratio exceeds 10\%. This trend is in contrast to the monotonically decreasing function reported in the literature, many of which used X-ray data to constrain the ERDF. 
\item The type 1 active fraction is $\sim$3\% for galaxies with $\mstar>10^{10}\msun$. Adopting that the type 1 AGNs constitute 7\%--20\% of all AGNs with $\mbh<10^8\msun$, we constrain the active fraction as 40\%--15\%. In the case of dwarf galaxies with $\mstar\sim10^8\msun$, the active fraction is $\sim$2\% even with a lower limit of the type 1 fraction of 7\%.
\item We constrain the black hole occupation fraction using the derived active fraction. For dwarf galaxies with $\mstar\sim10^{8.5}$--$10^9\msun$, we find that the occupation fraction is 50\%, which is consistent with the constraints based on the X-ray detections in the literature. 
\end{enumerate}

%%%%%%%%%%%%%%%%%%
% Acknowledgment %
%%%%%%%%%%%%%%%%%%

\begin{acknowledgments}
We thank the anonymous referee for constructive comments, which helped improve the quality of the manuscript substantially. H.C. would like to thank Jubee Sohn, Ena Choi, Shu Wang, Dongkok Kim, and Mankeun Jeong for the helpful discussions. This work has been supported by the Basic Science Research Program through the National Research Foundation of Korean Government (2021R1A2C3008486). J.-H.W would like to thank his collaborators for various inputs, despite the serious budget cut suddenly imposed by the government.
\end{acknowledgments}

\bibliography{bib.bib}

\end{document}